\documentclass[lettersize,journal]{IEEEtran}

\usepackage{amsmath,amsfonts,amssymb}
\usepackage{mathtools, cuted}
\usepackage{xcolor}
\usepackage{hyperref}

\usepackage{algpseudocode}
\usepackage{algorithm}
\usepackage{diagbox}
\usepackage{array}
\usepackage[caption=false,font=normalsize,labelfont=sf,textfont=sf]{subfig}
\usepackage{textcomp}
\usepackage{stfloats}
\usepackage{url}
\usepackage{verbatim}
\usepackage{graphicx}
\usepackage{subfig}
\usepackage{xcolor}
\usepackage{hyperref}
\usepackage{caption}
\usepackage{subcaption}
\usepackage[affil-it]{authblk}
\usepackage{diagbox}
\def\BibTeX{{\rm B\kern-.05em{\sc i\kern-.025em b}\kern-.08em
    T\kern-.1667em\lower.7ex\hbox{E}\kern-.125emX}}
\usepackage{balance}

\usepackage{csquotes}

\newcommand{\mk}{\langle k \rangle}
\newcommand{\adel}{\mathcal{A}_\delta}
\begin{document}
\title{Tree Proof-of-Position Algorithms}
\author{Aida Manzano Kharman, Pietro Ferraro, Homayoun Hamedmoghadam, Robert Shorten 
\thanks{Imperial College London, Dyson School of Design Engineering}
}

\markboth{Journal of \LaTeX\ Class Files,~Vol.~18, No.~9, September~2020}%
{How to Use the IEEEtran \LaTeX \ Templates}

\maketitle

\begin{abstract}
We present a novel class of proof-of-position algorithms: Tree-Proof-of-Position (T-PoP). This algorithm is decentralised, collaborative and can be computed in a privacy preserving manner, such that agents do not need to reveal their position publicly. We make no assumptions of honest behaviour in the system, and consider varying ways in which agents may misbehave. Our algorithm is therefore resilient to highly adversarial scenarios. This makes it suitable for a wide class of applications, namely those where trust in a centralised infrastructure may not be assumed, or high security risk scenarios. Our algorithm has a worst case quadratic runtime, making it suitable for hardware constrained IoT applications.
We also provide a mathematical model that allows us to analyse T-PoP's performance for varying operating conditions. We then simulate T-PoP's behaviour with a large number of agent-based simulations, which are in complete agreement with our mathematical model. T-PoP can achieve high levels of reliability and security by tuning its operating conditions, both in high and low density environments. Finally, we also present a mathematical model to probabilistically detect platooning attacks.  
\end{abstract}


\section{Introduction}

An emerging problem in many domains concerns the verification that an individual, or an object, is in the location that it claims to be in. Examples of situations where such a need arises range from images in conflict zones to vehicles who need to prove their location in order to access services. Proof-of-position algorithms refer to a suite of algorithmic tools that are designed to enable such objects to prove their location. An immediate question that arises is why not use GPS to determine one's location. GPS signals can be spoofed \cite{tippenhauer2011requirements}, \cite{manesh2019detection}.

The question of proving one's position is trivial to answer when a centralised and trustworthy infrastructure exists. An example such an infrastructure of this could, for example, be a network of surveillance cameras that are operated by a trusted government agency. In such situations it is relatively straightforward for an agent to verify their claimed location. However, such architectures impose a very strong trust assumption, and should the centralised network be compromised or malicious, one can no longer ensure that the location proof is trustworthy. Consequently, making such a trust assumption greatly restricts the practicality of any solution relying on it.  

Hence we advocate for the need for a proof-of-position algorithm that is \emph{decentralised}. 
However, decentralisation alone does not suffice. Information on an individual's position is sensitive. Revealing it may incur safety risks to the user. Thus, the proof-of-position algorithms must also be privacy preserving. Namely, one should be able to provide a verifiable proof of their position, without having to publicly reveal their position. In addition, for any solution to be applicable to real-world use cases, it must be designed to withstand adversarial environments and account for individuals that lie and misbehave. In the context of proof-of-position, there are many ways an agent could lie: they may lie about their own position, but also about the position of others. 
With these considerations in mind, our objective in this paper is to provide a proof-of-position algorithm that is decentralised; private; and robust to adversarial attacks.

\section{Related Work}
A preliminary version of our work was presented at \cite{tpop-cdc}. The following {\em Related Work} section closely resembles the related work in \cite{tpop-cdc}, given that, to the best of our knowledge, since the publication \cite{tpop-cdc}, the only new proof-of-position protocol published in the literature is \cite{blockchain-pol}.  

A number of works have appeared in the literature addressing the problem of proving one's position; see for example \cite{nosouhi2018sparse, alamleh2020cheat, javali2016alice, wu2020blockchain, amoretti2018blockchain}. These proposals are unsuitable for high risk scenarios we consider because they introduce trust assumptions that may not hold in adversarial environments. They also \textit{de facto} re-introduce a degree of centralisation in their systems. We proceed by outlining some of this prior work.   

An early example of a decentralized proof-of-location scheme, known as APPLAUS, was introduced in \cite{zhu2011toward}. APPLAUS addressed collusion attacks through graph clustering and computing a `betweeness' metric to evaluate node trustworthiness. In \cite{zhu2011applaus}, nodes weakly connected in the graph are considered less reliable. They propose a time-decayed weight function and determine node trustworthiness based on the ratio of approvals to neighbours. This latter contribution serves as a starting point for our work. Whilst offering a number of valuable contributions, their proposal relies on a central server that stores information on the number of agents at particular time and location to detect fraudulent proof generation. This information can be used to estimate
whether a prover lies about not finding enough peers or always
finding the same peer based on various statistical techniques. Conversely, the security and integrity of our algorithm does not rely assuming access to a trusted central server that stores said information. 

Another scheme, SHARP, presents a private proximity test without revealing actual locations to a server \cite{zheng2012sharp}. They introduce a secure handshake method without needing pre-shared secrets, ensuring a witness\footnote{An agent that  verifies that they  see another agent wishing to prove their position.} can only extract the session key if in they are indeed in the vicinity of the prover \footnote{An agent that wishes to prove their position.}. Security is ensured by requiring that location tags are un-forgeable. However, a coerced user could generate a valid location proof and relay it to a malicious user in a different location. This means the location proof is not non-repudiable \cite{paar2009understanding}.

Vouch+ \cite{boeira2019decentralized}, another decentralized approach, addresses high-speed and platooning scenarios. However, its security relies on the assumption that the selected proof provider is honest. We consider this assumption to be too strong. Indeed, our algorithm provides probabilistic guarantees of detecting that a prover is lying about their position. SPARSE \cite{nosouhi2018sparse} makes collusion difficult by preventing provers from choosing their own witnesses. Their solution provides integrity, un-forgeability and non-transferability. However, this is achieved through the use of a trusted verifier, and it is assumed that they will not publish users’ identity and their data. We again consider this assumption too strong.  

The authors in \cite{blockchain-pol} propose a proof-of-position protocol using blockchain agnostic smart contracts. A number of proposals have emerged to design proof-of-position schemes leveraging blockchain technologies, see \cite{amoretti2018blockchain}, \cite{wu2020blockchain}, \cite{nosouhi2020blockchain}, \cite{victor2018geofences}. Blockchain is known to centralise over time \cite{beikverdi2015trend}, and is highly resource intensive \cite{sedlmeir2020energy}. Our protocol does not rely on blockchain, thus avoiding the drawbacks inherent to its design.  

\textbf{Contributions with respect to prior-art:} In this paper we present a decentralised proof-of-position algorithm, T-PoP. It is decentralised because it does not rely on a trusted central server or authority to ensure the position proof is valid. It is privacy preserving, because the position proof may be computed without revealing the position of an agent publicly, and other participants may verify that the position proof is indeed valid. Indeed, we provide a proof of concept using Zama's fully homomorphic encryption library \cite{chillotti2020concrete}, where agents provide their encrypted position, and run T-PoP without having to decrypt it. It is available on our \href{https://github.com/aidamanzano/Tree-Proof-of-Position/blob/main/FHE.py}{GitHub} repository. Our algorithm is also robust to highly adversarial scenarios: we consider that agents may not only lie about their own position, but also about the position of other agents. We take this consideration into account to provide a probabilistic guarantee of an agent's position. 
We mathematically model the behavior of the algorithm, verify the model via agent-based simulations, and predict the algorithm's behavior in different scenarios analytically by the means of the verified model. This model provides designers with security and reliability guarantees of T-PoP's performance, given an expected distribution of malicious agents in the network. Thus, designers may use this model as opposed to conducting resource intensive agent based simulations to determine what operating conditions of our algorithm best suit their needs. Our algorithm runs in quadratic time (see section \ref{time-complexity}) and thus can be implemented in a wide range of IoT applications. T-PoP is a solution to prove one's location in a highly adversarial and mutually distrusting environment, where a central authority and trusted infrastructure cannot be relied upon. We demonstrate it performs well in both high and low density scenarios, given that its operating conditions may be tuned to provide increased security or reliability. 

\section{Components}
\noindent We proceed by introducing the components necessary to understand the functioning of T-PoP.\\

\noindent\textbf{Agents:} Any participant in the proof-of-position protocol is considered an agent, $a_i$, where $a_i \in A$, $|A| = N$ and $N$ is the total number of agents operating in the protocol. All agents are in a given \emph{state}, that determines if they are honest, and coerced, or not. For brevity, we omit the agent's index where possible.\\

\noindent\textbf{Environment:} We assume that agents willing to participate in the protocol are situated in in an environment, $E$, where $E \subseteq \mathbb{R}^2$. We note that this does not represent a limitation of the protocol and that any space with a suitable distance metric can be employed for the T-PoP protocol.\\

\noindent\textbf{Position:} All agents have a \emph{real} position, $p$, and a \emph{claimed} position, $\hat{p}$. Each agent can claim to be in a position different to their real one (i.e.: lie about their position) by committing to $\hat{p}$ such that $\hat{p} \neq p$.\\
\begin{subequations}
\begin{align}
    \hat{p} &\neq p \vee \hat{p} = p
\end{align}
\end{subequations}
In other words, the \emph{claimed} position may or may not be equal to the \emph{real} position. We say that $\hat{p}, p \in E$. \\

\noindent\textbf{Agent states:} All agents are in a given \emph{state}, which is comprised of three attributes: honesty, coercion and position. 
\begin{itemize}
    \item An agent is considered \emph{honest} if they claim to be in their real position. i.e.: $\hat{p} = p$, and \emph{dishonest} otherwise, where $\hat{p} \neq p$.
    \item All agents are in a real position. \emph{Honest} agents will claim to be in their real position, and \emph{dishonest} agents will claim to be elsewhere. 
    \item An agent is considered \emph{coerced} if they attest to seeing a \emph{dishonest} agent in their claimed (fake) position. In other words: a \emph{coerced} agent will claim to see agents that are not actually in its vicinity, if the latter are dishonest. \emph{Non-coerced} agents will only attest to seeing other nearby agents in their \emph{real} position.
    
\end{itemize}

\noindent\textbf{Prover:} An agent initiating the proof-of-position protocol, wishing to prove that they are in their \emph{claimed} position, is called a \emph{prover}.\\

\noindent\textbf{Range of sight:} Agents have a fixed, positive range of sight, $r_i$ with $r_i \geq 0$.\\

\noindent\textbf{Field of view:} Agent $a_i$'s field of view, $\overline{D_i}$, is defined as the closed disk area with centre $\hat{p_i} = (x_i, y_i)$ and radius $r_i$.
\begin{equation}
    \overline{D_i}= \{ (x, y) \in \mathbb{R}^2 : (x - x_i)^2 + (y - y_i)^2 \leq r_i^2 \}
\end{equation}
We say agent $a_j$ with position $\hat{p_j} = (x_j, y_j)$, is in $a_i$'s field of view if:
\[(x_j - x_i)^2 + (y_j - y_i)^2 \leq r_i^2\]
In other words, if the position of $a_j$ falls inside $\overline{D_i}$. \\

Note that this field of view is not restrictive of physical obstacles blocking immediate range of sight, rather, it is intended as a model for connectivity range.\newline

\noindent\textbf{Witness:} Provided $a_j$ is in $a_i$'s field of view, $a_j$ may be named as a \emph{witness} of $a_i$, depending on the \emph{states} of both agents. \\ 

\noindent\textbf{Approvals:} If a witness $a_j$ attests to seeing the agent that named it, $a_i$, the witness is said to \emph{approve} $a_i$. Approvals are always computed from both the agents' \emph{claimed} position.


\section{Overview of our protocol}

    
In this section, we outline a brief, high-level description of the protocol we propose. It is structured around three stages: \textbf{Commit}, Tree Building, and \textbf{Verification}. This protocol is initiated by an agent wishing to prove they are in a claimed position. In the initial stage, \textbf{Commit}, each agent constructs a cryptographic commitment of their claimed position, using a cryptographic commitment scheme\footnote{Many options are available: KZG commitment schemes \cite{kate2010polynomial}, Pedersen commitments, Merkle trees or Hashing \cite{damgaard1998commitment}. Some schemes also offer the desirable property of being additively and/or multiplicatively homomorphic \cite{frederiksen2015complexity}. The choice of commitment scheme will be driven by application specific needs.}. This commitment ensures the position claim cannot be changed \emph{a posteriori}, i.e., it is binding, and is also hiding, meaning the position need not be publicly revealed. Upon constructing the commitment, each agent then uploads their commitment onto a decentralised network. This can be a transaction in a DLT or uploaded to IPFS, a peer to peer file sharing network \cite{benet2014ipfs}. In this manner, immutability and verifiability are assured. 

Subsequently, the Tree Building stage follows. It may only be computed by an agent that has already committed to a position in the \textbf{Commit} stage. Said agent is considered the prover. The prover constructs a tree by naming other nearby agents, which are considered the prover's witnesses. Each node in the tree is an agent, where the prover is the root of the tree, and the witnesses are the leaves. The tree's dimensions are predetermined \emph{a priori}, namely the height and the branching factor. The protocol imposes strict constraints to prevent the naming the same witness twice in the same tree, aiming to mitigate deceitful practices (e.g., agents cyclically validating each other). 

The final stage of the protocol is the Verification stage. In it, the prover's tree of witnesses is assessed to verify the prover's position. The witnesses may attest to seeing the agent that named them, or not. If a witness does so, this is considered an approval. If sufficient approvals in the tree are amassed, the prover's position claim is considered valid, provided other necessary criteria are also met in the tree. 

\section{Implementation details of T-PoP}
We proceed by giving a more detailed description of the novel components we present in the T-PoP protocol. 

\begin{itemize}

    \item \emph{\bf Tree Building}:  Each agent who has already committed their position, then constructs a tree of depth $h \in \mathbb{N^+}$, incorporating the committed positions of agents, called \emph{witnesses}, at levels $d\in\{0,\ldots,h\}$. A specific agent ---which we denote as $g$---is the root of the resulting tree.  
    For every \emph{prover}, $g$, the tree is constructed as follows: 
     \begin{itemize}
         \item $g$ is is the root node at level 0.
         \item For each $d \in \{1,...,h\}$, each node at level $d-1$ will name $w_{d}$ other agents. An agent at level $d$ is an agent, $a_j$, that is in the range of sight of agent, $a_i$ at level $l-1$ (note that, if $\hat{p}_i \neq p_i$, and $a_i$ is lying about their position, it is possible that $a_j$ is not in the range of sight of $a_i$). $a_i$ is called the \emph{parent} of the $child$ $a_j$.
     \end{itemize}
     In practice, the root node, $g$, names $w_1$ witnesses who in turn  name $w_2$ witnesses and so on, until we reach depth $h$.  The number of witnesses per level, $n_l$, can therefore be computed recursively:
     \begin{equation}
         n_d = w_d n_{d-1},\;d=1,\ldots,h,
         \label{eq:nl}
     \end{equation}
    with $n_0 = 1$.  
    \item \emph{\bf Checks and Verification}: The Checks and Verification stages start by considering level $d = h - 1$ of the tree: 
    \begin{itemize}
        \item[1] Each witness at level $d+1$ states whether their parent at level $d$ is their neighbour or not (the child \emph{approves} the parent). If the answer is yes, and the child has not yet been named in the tree, this witness becomes a confirmed level $d$ witness. The total number of confirmed level $d$ witnesses is denoted as $D_d \leq n_d$, and the total number of witnesses that confirm a parent $b$ is denoted by $K_b \leq w_l$. 
        \item[2] If a witness has already been named before, this witness is removed, regardless of whether they confirm their parent or not. 
        \item[3] If $K_b<t  w_d$,  $t \in (0,1]$, parent $b$ is pruned from the tree. Here, $t$ is a parameter of T-PoP, called the {\em threshold}, which is used to regulate the security and reliability properties of the algorithm. 
        \item[4] If $D_d <  t  n_d$ then the algorithm interrupts and outputs that root $g$ is lying about their position. Otherwise, we move on to level $d-1$ and we repeat this process. Note that any parent removed by the previous step will not be included in this next iteration of T-PoP.
    \end{itemize}
\end{itemize}

T-PoP is therefore an algorithm depending on  a set of parameters, $\theta \equiv \{t, h, w_1, ..., w_h\}$.   The influence of these parameters on the performance of the algorithm will be explored in Section \ref{sec:sim}.  

\begin{figure}
    \centering
    \includegraphics[scale = 0.23]{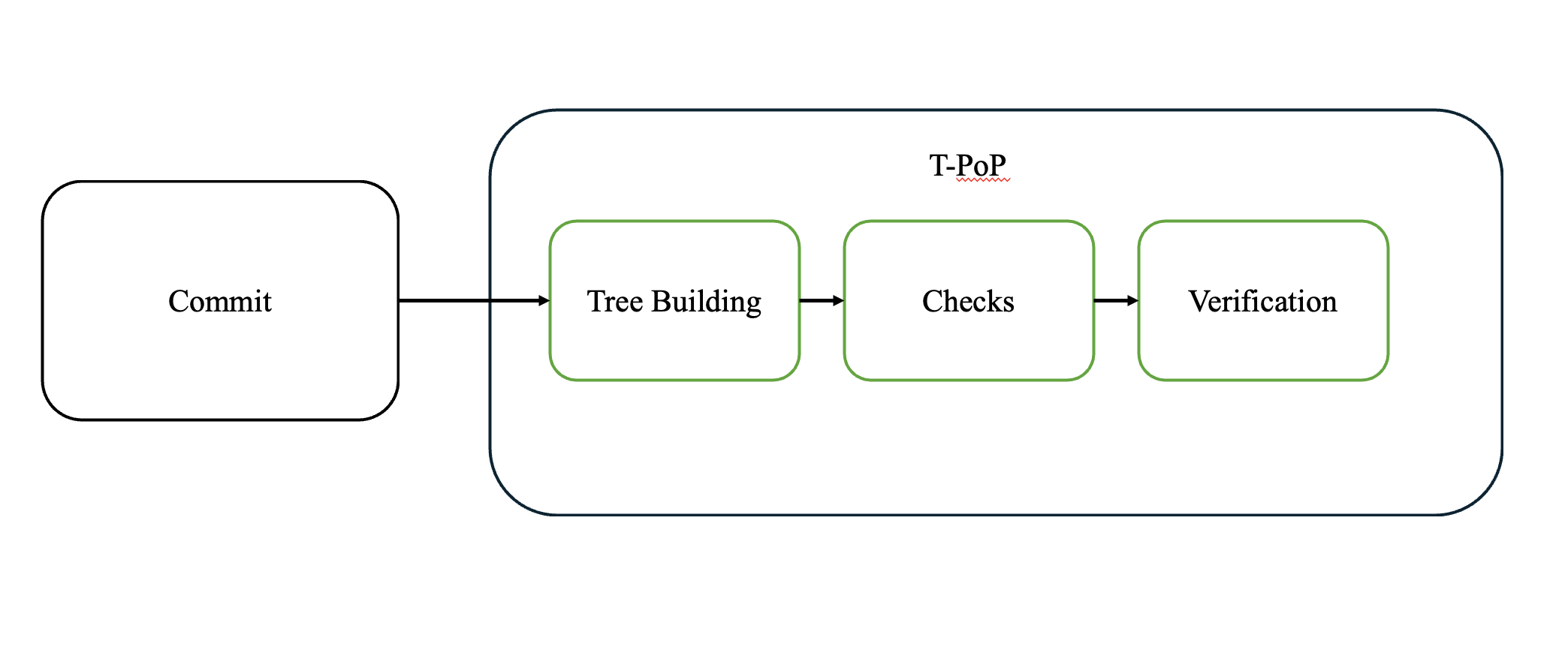}
    \caption{High-level Overview of the Proof of Position protocol.}
    \label{fig: TPoPArch}
\end{figure}

\begin{figure}
    \centering
    \includegraphics[scale = 0.3]{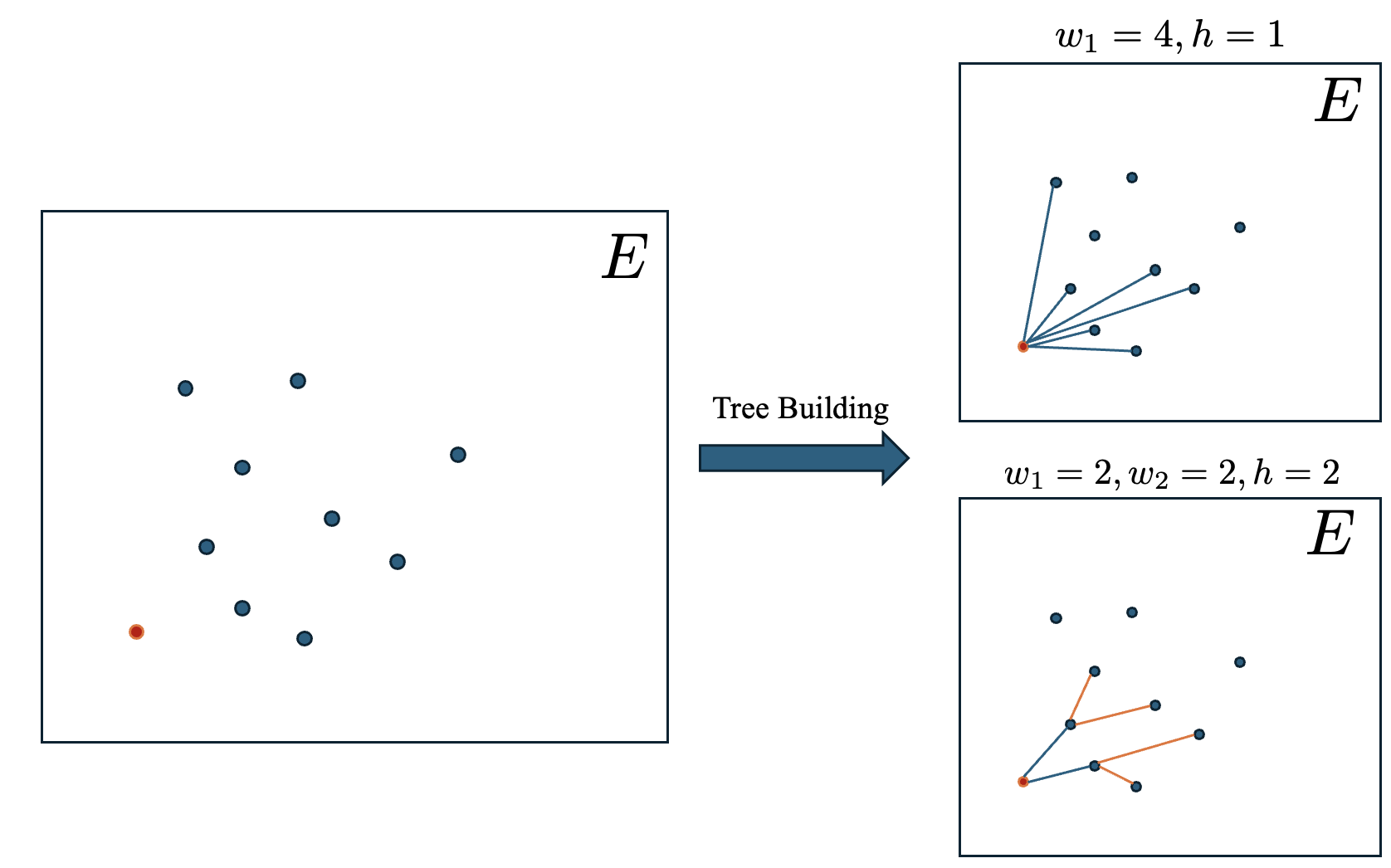}
    \caption{Tree building examples.  Agent $a$ commits their alleged position. The panel on the top right shows the construction of a tree for $h = 1$ and $w_1 = 4$, while the panel on the bottom right shows the construction of a tree for $h = 2, w_1 = 2, w_2 =2$.}
    \label{fig: T-PoPTree)}
\end{figure}

\begin{figure}
         \centering
         \includegraphics[width=0.49\textwidth]{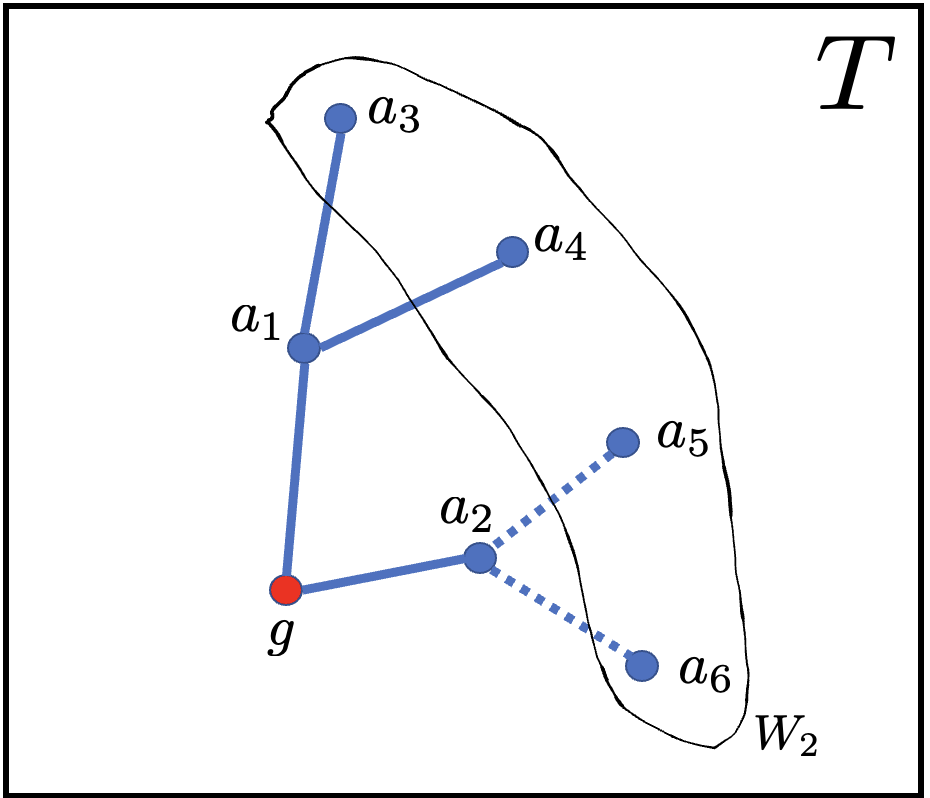}
         \caption{We start by evaluating the outer level of the tree and we evaluate the witnesses in $W_2$ (the circled set). Agents $a_5$ and $a_6$ do not confirm that they see agent $a_2$, even though $a_2$ is an honest agent. This leads to agent $a_2$ being eliminated from the tree.}
         \label{fig:example a}
     \end{figure}

\begin{figure}
         \centering
         \includegraphics[width=0.49\textwidth]{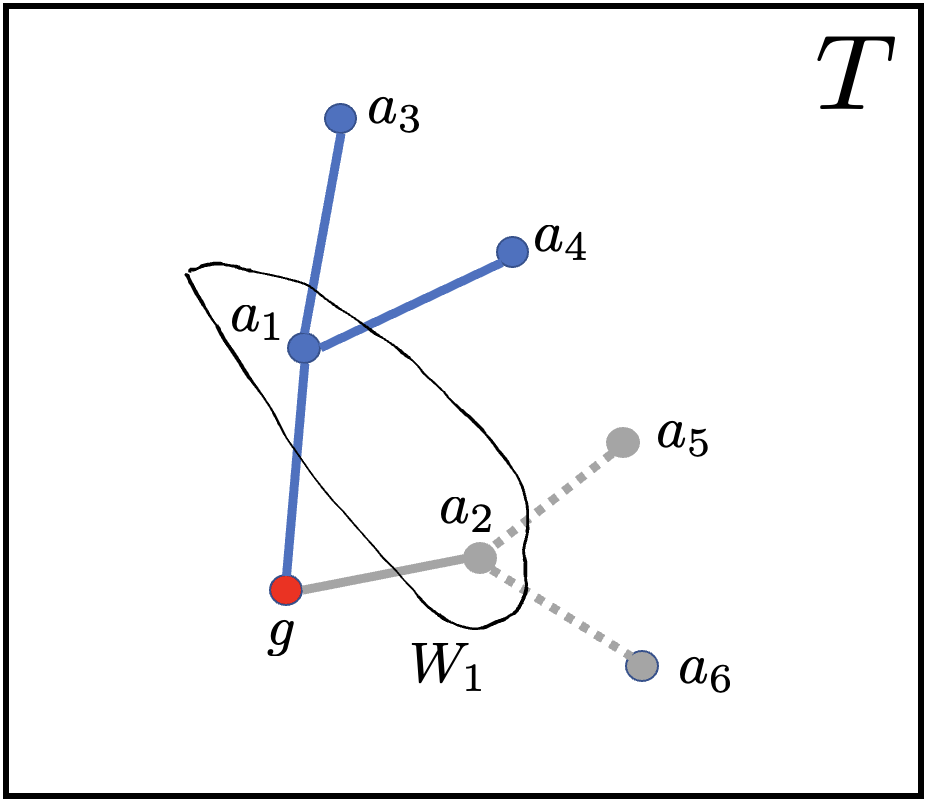}
         \caption{We go down one level, and now evaluate the witnesses in $W_1$ (the circled set). $a_2$ has been eliminated by the tree (shown in grey) and so only agent  $a_1$ is left.}
         \label{fig:example b}
     \end{figure}

{\bf Example:} Consider the  T-PoP example in Figures \ref{fig:example a} and \ref{fig:example b}, in which  $\theta = \{t =0.5, d = 2, w_1 = 2, w_2 = 2\}$, and so $n_1 = 2$ and $n_2 = 4$ (\ref{eq:nl}). Solid arrows mean that a witness approves their parent and dotted lines mean that a witness does not approve their parent. Agents $a_5$ and $a_6$ are  dishonest agents, so that their committed positions, $\hat{p}_5$ and $\hat{p}_6$, are different from their true positions. However, agent $a_2$  does not know this, it saw those cars next to it and it picked $a_5$ and $a_6$ as witnesses. So, $a_5$ and $a_6$ do not confirm that $a_2$ is a neighbour of theirs, whereas $a_3$ and $a_4$ confirm that $a_1$ is a neighbour of theirs. In line with point 3 of \emph{Checks and Verification} (above), agent $a_1$ has enough confirmed witnesses ($K_{a_1} = 2 \geq t\times w_2 =  0.5 \times 2$) and stays in the tree, while agent $a_2$ does not have enough confirmed witnesses ($K_{a_2} = 0 < 0.5 \times 2$), and so $a_2$ is removed from the tree. However, since the total number of confirmed witnesses at level 2 is $D_2 = 2 \geq t \times n_2  = 0.5 \times 4 $, T-PoP does not stop for $g$, and we move to level 1. At level 1, $a_2$ has been removed but $a_1$ confirms that $g$ is its neighbour. As per points 2 and 3 of \emph{Checks and Verification}, the final output of T-PoP is that $g$ is {\em truthful\/} about their position. As can be seen in the example above, $t$ is critical in determining the output of T-PoP.  For instance, if $t=1$, then $M_2 = 2 < t \times n_2  = 1 \times 4 = 4$, causing T-PoP to stop at point 4 of \emph{Checks and Verification}, and returning an output of {\em untruthful\/}  for $g$. 
 
     \hfill

T-PoP's practical implementations will vary depending on the application. It may be used in by vehicles to prove their position within a city, or it could be implemented in a smartphone, operated by individuals wishing to prove their location. It is also worth noting that T-PoP is not a highly resource intensive algorithm to execute, and thus the hardware constraints are not a critical concern. It is comprised of two algorithms: \textbf{Verification} and \textbf{Checks}. These are further defined in \ref{verification-algorithm} and \ref{checks algorithm} accordingly. \textbf{Verification} is a (reverse) breadth-first-search algorithm, meaning its runtime in the worst case scenario is $O(|V| + |E|)$, where $V$ is the number of nodes and $E$ the number of edges in the tree data structure \cite{cormen2022introduction}. The \textbf{Verification} algorithm calls the \textbf{Checks} algorithm at each node, and the \textbf{Checks} algorithm has a time complexity of $O(|V|)$. So T-PoP's time complexity is $O(|V|^2 + |E||V|)$. \label{time-complexity}

Of significant importance are the identity management solution, the commitment scheme selected, the communication security protocol and the notarising of the position proofs. 
Subsequently, we proceed with providing some possible options, but we note that the final decision must be made considering the context specific needs: conflict-zones will have different security and privacy needs to smart cities. 

Identity management is trivial using a centralised solution. T-PoP is a decentralised, peer to peer algorithm, so should a centralised trust assumption not be desirable, we suggest other alternatives: Proof of Humanity (PoH) is a decentralised project \textit{`to provide a social identity verification system for humans on Ethereum'} \cite{PoH21}. Users cannot submit duplicate identities, fake or deceased identities may be challenged and removed and identities are verified by other verified and registered users. The drawback of this mechanism is that the identity is (naturally) public. For reporting purposes this is not an issue: paired with a commitment scheme, the reporter's position remains hidden, but the proof of humanity shows the reporter (prover) is a real and verified individual. PoH is built on the Ethereum blockchain, thus incurring transaction costs to the user wishing to obtain a proof of humanity, however, this is a one time cost, as users will not obtain more than one proof of humanity.

Regarding securing communication between agents in T-PoP, we advise using established protocols such as TLS 1.3. The protocol used should guarantee the following properties are met:
\begin{enumerate}
    \item Confidentiality: Information is kept secret from all but authorized parties.
    \item Integrity: Messages have not been modified in transit.
    \item Message Authentication: The sender of the message is authentic.
    \item Non-repudiation: The sender of the message cannot deny the creation of the message. 
    \cite{paar2009understanding}
\end{enumerate}

Finally, upon gaining a proof-of-position, this may be easily notarised in a DLT or IPFS, along with the prover's position commitment and prover's tree, such that all other agents may verify that with the same inputs, they obtain the same T-PoP result. This must be accompanied with the prover and the witnesses' unique signature to prevent forgery.

\section{TPoP Protocol} \label{tpop-protocol}
In this section we provide an overview of T-PoP from an algorithmic perspective and we introduce some notation that is used in the remainder of the paper. 

Accordingly, there are three main algorithms that T-PoP calls:

\begin{itemize}
    \item \emph{Tree Building:} the process of building the tree of witnesses.
    \item \emph{Checks:} the process of verifying that the tree that has been built satisfies certain criteria, listed below.

    \item \emph{Verification:} the process of assessing the tree and deciding whether or not to confirm the prover's position.
\end{itemize}

\subsection{Tree Building}
A \emph{prover's} aim is to amass sufficient \emph{approvals} to support their position claim. Indeed, the prover provides a tree data structure that summarises their received approvals and the dependencies between them. 
The number of nodes and edges present in the tree are evaluated to determine the likelihood of the prover's claim being true. 
We proceed by outlining the algorithm the prover follows when constructing the tree structure of approvals.




\noindent\textbf{Terminology:} Each node in the tree represents an agent. The root of the tree is the prover. In a pair of nodes connected by a directed edge, the node where the edge originates is the \emph{parent}, and the node where the edge terminates is the \emph{child}. Each edge represents an \emph{approval}. Namely, an edge from parent $a_i$ to child $a_j$ encodes $a_j$'s approval of $a_i$. The tree must be of a specific dimension. Precisely, the variables are:
\begin{itemize}
    \item \textbf{height}: the size of the longest path from the root to the leaves of the tree, $h$.
    \item \textbf{branching factor}: the number of children, $w_d$, that a parent must have at a given depth level, $d$. For brevity, if the branching factor is constant across levels, we omit the index and simply denote it as $w$.
\end{itemize}
\begin{algorithm}
\caption{Tree Construction} 
	\begin{algorithmic}[1]
            \Require{Prover $a_i$, Height $h$, Branching Factor $w_1, ..., w_d$}
            \State{Initialise $a_i$ as the root of the tree $g$ and as a parent of depth level $1$}
                \For{$d = 1, ..., h - 1$}
                    \For{each parent $a_i$ at depth level $d$}{}{}
                        \State
                        \parbox[t]{\dimexpr\linewidth-\algorithmicindent}{$a_i$ names $w_{d+1}$ children among its neighbours}
                        \State
                        \parbox[t]{\dimexpr0.91\linewidth-\algorithmicindent}{All the named children are added as nodes of $G$ at depth level $d+1$, with $a_j$ as their parent node.}
                    \EndFor
                \EndFor
            \State{{\bf return} G}
	\end{algorithmic} 
        \label{tree algorithm}
\end{algorithm}

\subsection{Checks Algorithm}
This algorithm is called within the \textbf{Verification} algorithm, and its purpose is to assess, for each parent's sub-tree, whether it satisfies certain criteria. If so, the parent node is considered honest and therefore remains in the prover's tree. Otherwise, it and its branch are pruned from the prover's tree.
The criteria being checked for each sub-tree defined in algorithm \ref{checks algorithm}, and are the following:

\begin{enumerate}
    \item each child \emph{approves} their parent (Line 1), \label{criteria1}
    \item the parent has named sufficient children (Line 2),\label{criteria2}
    \item no child has been already named in the tree by another parent (Line 3),\label{criteria3}
    \item no parent names the same child twice (Line 4).\label{criteria4}
\end{enumerate}
Indeed, criterion 3 and 4 can be summarised by ensuring each agent in the tree is unique. The threshold parameter, $t$, is a percentage that determines the minimum number of children that a parent must have. A parent remains in the tree \emph{iff} $|parent's\ children| \geq w_d \cdot t$.
\begin{algorithm}
\caption{Checks} 
	\begin{algorithmic}[1]
            \Require{Child $a_j$, Parent $a_i$, Named Agents $A$, Branching Factor $w_d$, threshold $t$}
            \If{$a_j$ \emph{approves} $a_i$ \\
            \textbf{and} $\#\text{[$a_i$'s children]}\geq w_d \cdot t$\\
            \textbf{and} $a_i$ not in $A$\\
            \textbf{and} $\#set$(\text{[$a_i$'s children]}) = $\#$\text{[$a_i$'s children]}}
            \State{\textbf{Add} $a_j$ to $A$}
            \State{{\bf return} True}
            \Else
            \State{{\bf return} False}
            \EndIf

	\end{algorithmic} 
        \label{checks algorithm}
\end{algorithm}

\subsection{Verification}
Starting at the lowest depth level, for each parent in that level, \textbf{Verification} calls the \textbf{Checks} algorithm for all of its children. If successful, the parent and its branch remains in the tree, otherwise, it is pruned. It then moves up a level, and performs the same operations, only considering the non-pruned nodes. The algorithm terminates at the root of the tree, where it evaluates if the tree has sufficient nodes after pruning. If so, the \emph{prover} is considered honest, otherwise it is considered dishonest. 
Using the prover's tree and the \textbf{Checks} function, \textbf{Verification} calls \textbf{Checks} in a reverse breadth-first-search manner. 
The algorithm returns True if the \emph{prover} meets the required criteria and False otherwise.

\begin{algorithm}
\caption{Verification} 
	\begin{algorithmic}[1]
            \Require{Tree $G$, threshold $t$, Height $h$, Branching Factor $w_1, ..., w_d$, threshold $t$}
            \State{Initialise Named Agents to $A = $ []}
            \State{Initialise $G$'s nodes to \emph{not pruned}}
            \State{Initialise Depth Level Approvals to $D_1, ..., D_{d-1}$}
            \For{$d = 1, ..., h - 1$}
                \State{Depth Level Approval: $D_d = 0$}
                    \For{each parent in at depth level $d$}
                    \State{$Parent\ Approvals = 0$}
                        \For{each parent's child}{}{}
                            \If{child is \emph{not pruned}}
                                \If{\textbf{Checks}(child, parent, $A$, $w_d$, $t$) is True}
                                \State{$Parent\ Approvals\ += 1$}
                                
                                \EndIf
                            \EndIf
                        
                        \EndFor
                    \If{$Parent\ Approvals < w_d \cdot t$}
                        \State{parent is \emph{pruned}}
                    \Else
                    \State{$D_d\ += 1$}
                    \EndIf
                    \EndFor
                \If{$D_d < w_d \cdot t$}
                    \State{{\bf return} False}
                \Else
                    \State{{\bf return} True}
                \EndIf
            \EndFor

	\end{algorithmic} 
        \label{verification-algorithm}
\end{algorithm}

\section{Mathematical Modelling of T-PoP} \label{sec: math model approval}
We proceed with mathematically characterising the functioning of T-PoP. This mathematical model can be used to select the most suitable operating conditions of T-PoP, $\theta = (h, w, t)$, provided the designers select a desired performance level they wish to achieve in their application. 

As discussed in Section \ref{tpop-protocol}, the heart of T-PoP is the \textbf{Checks} algorithm, which is recursively called in the \textbf{Verification} algorithm. The \textbf{Checks} algorithm assesses whether the criteria necessary to confirm an agent's position claim are satisfied. These criteria are the following: 
\begin{enumerate}
    \item each child \emph{approves} their parent, 
    \item each parent has named sufficient children,
    \item all agents are unique in the tree
\end{enumerate}

We hypothesize that it suffices to compute the probability that each of the criterion above are met to determine the probability that an agent will receive a valid proof-of-position. We proceed by computing the individual probabilities of each criterion and finally we combine them into one model. Consequently, the probability that an agent $a_i$ with state $s_u$ will receive a proof-of-position through T-PoP with operating conditions $\theta$, is equal to the product of the probability of each criterion above being satisfied:
\begin{equation}
    P(\text{TPoP}(a) = \text{True}| s_u,\theta) = \prod_{j=1}^3 P(\text{Criteria $j$ $=$ True} |  s_u,\theta) \nonumber
\end{equation}

Finally, in order to analyse T-PoP's performance, we consider two metrics: Secutity and Reliability. 

\begin{itemize}

    \item {\em \bf True Negatives or Security, $TN$, \/} is a conditional probability  quantifying the ability of the algorithm to detect malicious agents. Specifically, it is the true-negative conditional probability, which, under stationarity assumptions, is independent of $i\in \{1, \ldots,|A|\}$:
    \[
TN \equiv \Pr[\text{TPoP}(a) = \text{False} | a \in \overline{H} ]
\]

   \item {\em \bf True Positives or Reliability, $TP$,\/} is a conditional probability  quantifying the ability for the algorithm to detect honest agents. Specifically, it is the true-positive conditional probability.
    Once again, under stationarity assumptions: 
 \[
TP \equiv\Pr[\text{TPoP}(a) = \text{True} | a \in H ]
\]

\end{itemize}

In what follows, we derive a mathematical expression for $P(\text{Criteria $j$ $=$ True} |  s_u,\theta), j\in\{1,2,3\}$, and subsequently demonstrate that our model faithfully represents T-PoP's behaviour. We do so by conducting extensive agent-based simulations, and find that the results obtained are in complete agreement with those of the mathematical model we provide.  

The rationale behind our analysis is to understand the performance of the algorithm when multiple agents collude together to try and, either fake their own position (dishonest agents) or attest that they see another agent in their fake position (coherced agents). 
We first begin by formalising the agent states mathematically. This allows us to construct a probability model to determine the likelihood that a prover will obtain sufficient approvals in their tree.

\subsection{Agent States} \label{agent-states}
All agents have three attributes: honesty, $\alpha$, coercion, $\beta$ and a claimed position, $\gamma$. These attributes are binary, taking either a \textbf{True} or \textbf{False} value. We say an agent $a_i$ has an honesty attribute $\alpha_i$, coercion attribute $\beta_i$ and claimed position $\gamma_i$, where $a_i \in A$, recalling that $A$ is the set of all agents.
\begin{itemize}
    \item \emph{Honesty:} An agent $a_i$ is either \emph{honest}, $\alpha_i = h$ or \emph{dishonest}, $\alpha_i = \bar{h}$. An \emph{honest} agent claims to be in their \emph{real} position, whereas a \emph{dishonest} agent claims to be in a position that is different to their \emph{real} position. We call the  set of \emph{honest} agents $H=\{a_i\in A : \hat{p}_i = p_i\}$ and the set of \emph{dishonest agents} $\overline{H} = \{a_i \in A: \hat{p}_i \neq p_i\}$.\\
    
    \item \emph{Coercion:} An agent $a_i$ is either \emph{coerced}, $\gamma_i = c$ or \emph{non-coerced}, $\gamma_i = \bar{c}$. When approving other agents, a \emph{coerced} agent will approve other agents in their \emph{claimed} position (regardless of whether this \emph{claimed} position is equal to their \emph{real} position or not). A \emph{non-coerced} agent will only approve agents in their \emph{real} position.
    We call the  set of \emph{coerced} agents $C=\{a_i\in A : \beta_i = c\}$ and the set of \emph{non-agents} $\overline{C} = \{a_i \in A: \beta_i = \overline{c}\}$.\\
    
    \item \emph{Position:} An agent $a_i$ claim to be in a position. If the agent is \emph{honest}, their \emph{claimed} position is equal to their \emph{real} position $p$. As such, we say, $\gamma_i = t$. If the agent is \emph{dishonest}, their \emph{claimed} position $\hat{p}$ is different to their \emph{real} position, $p$, and therefore, $\gamma_i = \bar{t}$.
    We call the  set of agents that claim to be in their real position $T=\{a_i\in A : \gamma_i = t\}$ and the set of agents that claim to be elsewhere $\overline{T} = \{a_i \in A: \gamma_i = \overline{t}\}$.\\
\end{itemize}

We denote all three attributes of an agent $a_i$ as follows:
\begin{subequations}
    \begin{align}
        \alpha_i &\in \{h, \overline{h}\} \\
        \beta_i &\in \{c, \overline{c}\} \\
        \gamma_i &\in \{t, \overline{t}\}
    \end{align}
\end{subequations}

\begin{figure}[!t]
\centering
\includegraphics[width=0.4\textwidth]{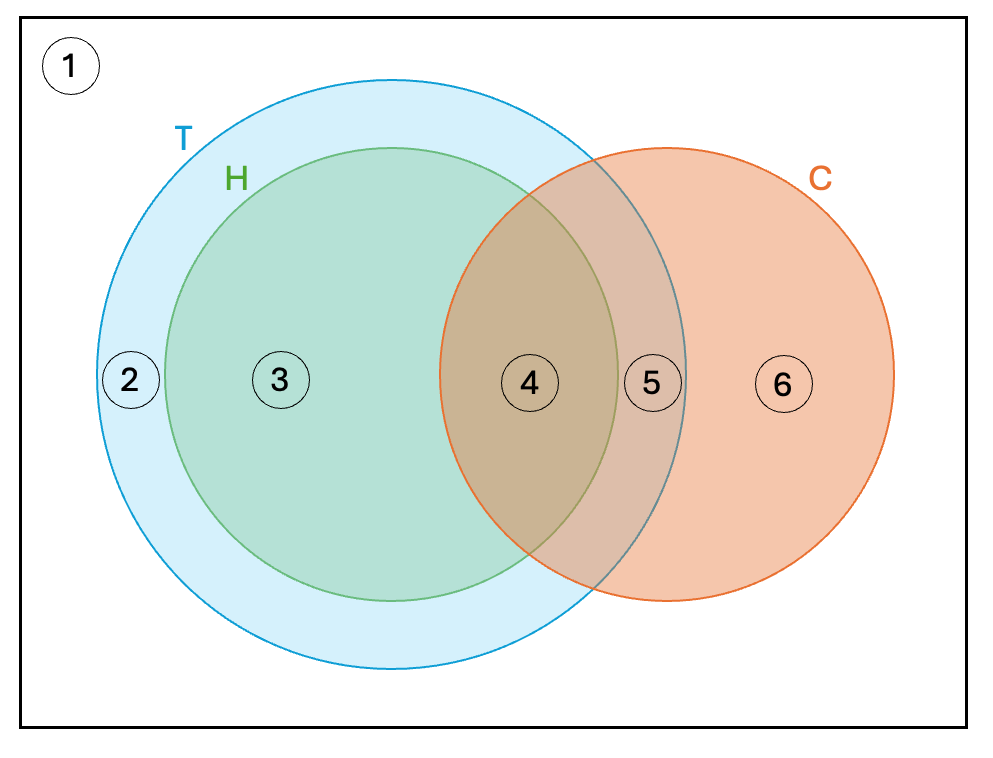}
\caption{Venn Diagram Representing Possible Agent State Combinations}
\label{venn-diagram}
\end{figure}
We depict the agents having each attribute (honesty, coercion and position) as a set in Figure \ref{venn-diagram}. Each numbered region in Figure \ref{venn-diagram}, $u$, where $u \in \{1, 2, 3, 4, 5, 6\}$ corresponds to a possible set of attribute combinations, $S_u$, that an agent may possess. For example, if $a_i \in S_2$ then $a_i$ belongs to the set of all agents that are dishonest, $\overline{H}$, the set of all agents that are not coerced, $\overline{C}$, and the set of all agents that claim to be in their true position, $T$. 
For ease of exposition, we define the \emph{state} of an agent $a_i$, $x(a_i)$, to be the set of attributes an agent has:
\begin{equation}
    x(a_i) = \{\alpha_i, \beta_i, \gamma_i\}
\end{equation}

where $x(a_i)$ may be any of the following permutations:
\begin{subequations}
    \begin{align}
        s_1 = \{\overline{h}, \overline{c}, \overline{t}\} \\
        s_2 = \{\overline{h}, \overline{c}, t\} \\
        s_3 = \{{h}, \overline{c}, t\} \\
        s_4 = \{{h}, {c}, {t}\} \\
        s_5 = \{\overline{h}, {c}, \overline{t}\} \\
        s_6 = \{\overline{h}, {c}, {t}\}
    \end{align}
\end{subequations}

In short, we say that an agent $a_i$ has a state $x(a_i)$, and that it belongs to the set of agents $S_u$, ($a_i \in S_u$), where $i \in |A|$, $u \in \{1,2,3,4,5,6\}$ and
\begin{subequations}
    \begin{align}
        S_1 = \{\overline{H} \cap \overline{C} \cap \overline{T}\}\\
        S_2 = \{\overline{H} \cap \overline{C} \cap T\}\\
        S_3 = \{H \cap \overline{C} \cap T\}\\
        S_4 = \{H \cap C \cap T\}\\
        S_5 = \{\overline{H} \cap C \cap \overline{T}\}\\
        S_6 = \{\overline{H}\cap C \cap T\}
    \end{align}
\end{subequations}

\begin{subequations}\label{P(s_u)}
\begin{align}
    P(a_i \in S_u) &= P(x(a_i) = s_u) \\
    P(x(a_i) = s_u) &= P(\alpha_i) P(\beta_i)
    P(\gamma_i | \alpha_i)
   \end{align}
\end{subequations}

Formally, we assume that variables $\alpha$ and $\beta$ are binary random variables, independent of each other. These attributes represent the probability of an agent being honest and coerced respectively. We can therefore define the probability of an agent being honest (i.e.: $P(\alpha_i = h)$) as $p_h$, and being coerced (i.e.: $P(\beta_i = c)$) as $p_c$, where $p_h, p_c\in[0,1]$. Similarly, $1-p_h$ and $1-p_c$ are the probabilities of an agent being dishonest and non-coerced respectively. Formally

\begin{subequations}
    \begin{align}
        p_h = P(x(a_i) = s_u, \  \forall u \in \{3, 4\}) \\
        p_c = P(x(a_i) = s_u, \  \forall u \in \{4, 5, 6\})
    \end{align}
\end{subequations}
where $p_h, p_c \in[0,1]$. 

\begin{subequations}
    \begin{align}
    \label{eq: ph}
        &p_h = P(\alpha_i = h) \\ 
        \label{eq: pc}
        &p_c = P(\beta_i = c) \\ 
        &1-p_h = P(\alpha_i = \overline{h}) \\
        &1-p_c = P(\beta_i = \overline{c})
    \end{align}
\end{subequations}

Next, we wish to compute the following probabilities: $P(\alpha_i = h)$, $P(\beta_i = c)$ and $P(\gamma_i = t)$ where:

   \[P(\alpha_i = h) = 1 - P(\alpha_i = \overline{h})\]
    \[P(\beta_i = c) = 1 - P(\beta_i = \overline{c})\]
    \[P(\gamma_i = t)= 1 - P(\gamma_i = \overline{t})\]

This is not the case with $\gamma$. Attributes $\gamma$ and $\alpha$ are not independent. Indeed, by definition: 
\begin{subequations} \label{eq: gamma honest}
    \begin{align}
        & P(\gamma_i = t |\alpha_i = h) = 1 \\
        & P(\gamma_i = \overline{t} |\alpha_i = h) = 0
    \end{align}
\end{subequations} In practical terms: an honest agent will always claim to be in their real position.
If $P(\alpha_i = \overline{h})$, finding $P(\gamma_i = t)$ is slightly more convoluted. This is because when a dishonest agent is sampled in a tree, one cannot, in theory, distinguish if its claimed position is its real position or not. Indeed, that is the purpose of T-PoP: to determine whether an agent is claiming to be in their real position. 

\emph{Remark:} When a dishonest agent is named in a tree, it could be that the dishonest agent is being observed in their real position, or that it has somehow tricked other nearby agents into pretending it is in their vicinity, when in fact it is elsewhere (by sending a fake signal or through any other possible attack vector). Because of this, we consider dishonest agents to effectively exist in two places: their real position, and their claimed position, which is fake.

\begin{figure}

  \centering
  \subfloat[1][Dishonest parent]{\includegraphics[width=0.2\textwidth]{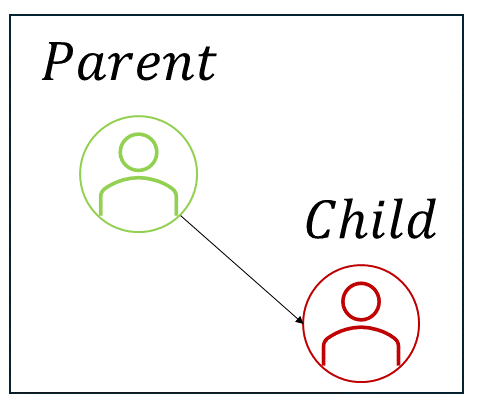}}\label{fig:dishonest parent}\\
  \subfloat[2][Dishonest child]{\includegraphics[width=0.2\textwidth]{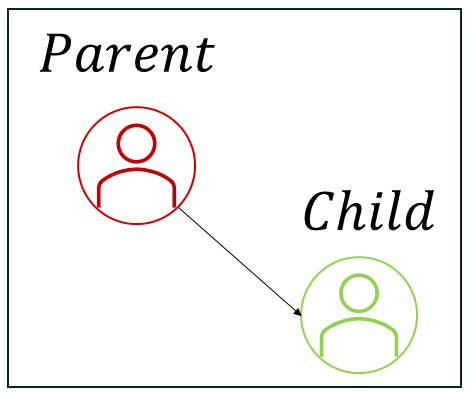}} \label{fig:dishonest child}\\
  \caption{Possible scenarios in which a dishonest agent may appear in a prover tree.} \label{fig:dishonest agents}
\end{figure}

To determine $P(\gamma_i = t)$, we must therefore consider the two possible cases in which a dishonest agent may appear in a tree: as a parent or as a child. This is depicted in Figure \ref{fig:dishonest agents}. 

Let us begin with the first case, shown in Figure \ref{fig:dishonest agents}a. If the dishonest agent $a_i$, is the parent in the tree:
\begin{subequations} \label{eq: gamma dishonest parent}
    \begin{align}
        & P(\gamma_i = t |\alpha_i = \overline{h}) = 0 \\
        & P(\gamma_i = \overline{t}| \alpha_i = \overline{h}) = 1 
    \end{align}
\end{subequations}

This is because the dishonest agent would never construct a tree for their real position: doing so would expose their lie. 

In the second case, as depicted in Figure \ref{fig:dishonest agents}b, if the dishonest agent is a child in the tree, then  $P(\gamma_i = t) = 1$ if their parent is \emph{non-coerced}, and  $P(\gamma_i = t) = 0$ if their parent is \emph{coerced}. 
Formally, for a dishonest child $a_i$, with a parent $a_j$:
\begin{subequations}\label{eq: gamma dishonest child}
    \begin{align}
     &P(\gamma_i = t | \alpha_i = \overline{h}, \beta_j = c) = 0 \\
     & P(\gamma_i = \overline{t}| \alpha_i = \overline{h}, \beta_j = c) = 1 \\
     & P(\gamma_i = t| \alpha_i = \overline{h}, \beta_j = \overline{c}) = 1 \\
     & P(\gamma_i = \overline{t} | \alpha_i = \overline{h}, \beta_j = \overline{c}) = 0 
    \end{align}
\end{subequations}

\emph{Remark:} Recall that a \emph{coerced} agent is one that covers up for other agents that lie about their position. This means that if they happen to name a dishonest agent as their child, they will name the dishonest agent in their fake position. The reader may consider the state of coercion as an agent being bribed, attacked or threatened to co-operate with another agent that is lying about their own position.

\subsection{Condition 1: does a child approve their parent}
Let us proceed by formalising the most crucial criterion: whether a child will approve the parent that named them.
\begin{figure}
\centering
\includegraphics[width=0.4\textwidth]{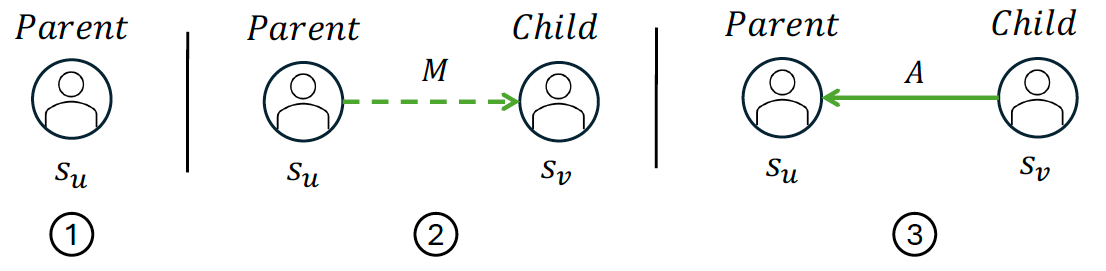}
\caption{Process of Parent to Child Approval}
\label{fig-checks1}
\end{figure}
There are three stages in the process of an approval, as depicted in Figure \ref{fig-checks1}. 

\subsubsection{Initial Parent State}
First, consider the probability of the parent being in state $s_u$ in equation \ref{P(s_u)}. The vector in equation \ref{eq-initial state} summarises the probability of the parent being in any of the initial states.
\begin{equation}
      I_1 = [P(s_1), P(s_2), P(s_3), P(s_4), P(s_5), P(s_6)]^T \label{eq-initial state}
\end{equation}
where $P(s_u)$ is short for $P(\text{parent} \in S_u)$  with $\sum_{k=1}^6 P(s_k) = 1$.

\subsubsection{Picking a child, given a parent}
In the next step, we determine the probability of picking a child in a given state, provided the parent is in another state. Let us denote the child as agent \textbf{c} and the parent as agent \textbf{p}. We then summarise the probability that a parent \textbf{p} in a state $s_u$ will select a child \textbf{c} in state $s_v$ in Table \ref{neighbours-logic}. 

\begin{table*}[h]
    \centering
    \def\arraystretch{1.5}
    \begin{tabular}{|c|c|c|c|c|c|c|}
        \hline
         \diagbox[width=8em]{Parent}{Child} & $\overline{H} \cap \overline{C} \cap \overline{T}$ & $\overline{H} \cap \overline{C} \cap T$ & $H \cap \overline{C} \cap T$ & $H \cap C \cap T$ & $\overline{H} \cap C \cap \overline{T}$ & $\overline{H} \cap C \cap T$\\
        \hline
        $\overline{H} \cap \overline{C} \cap \overline{T}$  & 0 & $m_{1,2}$ & $m_{1,3}$ & $m_{1,4}$ & 0 & $m_{1,5}$ \\
        \hline
        $\overline{H} \cap \overline{C} \cap T$ & 0 & 0 & 0 & 0 & 0 & 0 \\
        \hline
        $H \cap \overline{C} \cap T$ & 0 & $m_{3,2}$ & $m_{3,3}$ & $m_{3,4}$ & 0 & $m_{3,6}$ \\
        \hline
        $H \cap C \cap T$  & $m_{4,1}$ & 0 & $m_{4,3}$ & $m_{4,4}$ & $m_{4,5}$ & 0 \\
        \hline
        $\overline{H} \cap C \cap \overline{T}$ & $m_{5,1}$ & 0 & $m_{5,3}$ & $m_{5,4}$ & $m_{5,5}$ & 0 \\
        \hline
        $\overline{H} \cap C \cap T$ & 0 & 0 & 0 & 0 & 0 & 0 \\
        \hline
    \end{tabular} 
     
    \caption{Parent to Child Witness Selections}
    \label{neighbours-logic}
\end{table*} 

\emph{Note:} Certain parent to child combinations are not possible by defintion. This is shown by the entries of the Table \ref{neighbours-logic} that are $0$. For example, a \emph{non-coerced} parent will never name a \emph{dishonest} child in their \emph{claimed} (fake) position. This is because \emph{non-coerced} agents do not `see' agents in a their fake positions.

We summarise the information in Table \ref{neighbours-logic} as a matrix:
\begin{equation}
M = 
\begin{bmatrix} 
m_{1,1} & m_{1,2} & m_{1,3} & m_{1,4} & m_{1,5} & m_{1,6} \\
m_{2,1} & m_{2,2} & m_{2,3} & m_{2,4} & m_{2,5} & m_{2,6} \\
m_{3,1} & m_{3,2} & m_{3,3} & m_{3,4} & m_{3,5} & m_{3,6} \\
m_{4,1} & m_{4,2} & m_{4,3} & m_{4,4} & m_{4,5} & m_{4,6} \\
m_{5,1} & m_{5,2} & m_{5,3} & m_{5,4} & m_{5,5} & m_{5,6} \\
m_{6,1} & m_{6,2} & m_{6,3} & m_{6,4} & m_{6,5} & m_{6,6} \\
\end{bmatrix}\label{M}
\end{equation}

where the entry $m_{u,v}$ is defined as:
\begin{equation}
    m_{u,v} = P[x(\textbf{c}) = s_v) | (x(\textbf{p}) = s_u)]
\end{equation}
 and
 \[x(\textbf{c}) = s_v = \{\alpha_{\textbf{c}}, \beta_{\textbf{c}}, \gamma_{\textbf{c}}\} \]
\[x(\textbf{p}) = s_u = \{\alpha_{\textbf{p}}, \beta_{\textbf{p}}, \gamma_{\textbf{p}}\} \]
To compute $m_{u,v}$ it suffices to evaluate:
\begin{equation} \label{eq-m_ij}
    m_{u,v} = P(\alpha_\textbf{c})P(\beta_\textbf{c}) P(\gamma_\textbf{c})
\end{equation}

$P(\alpha_\textbf{c})$ and $P(\beta_\textbf{c})$ are known through equations \ref{eq: ph} and \ref{eq: pc} respectively. To evaluate $P(\gamma_\textbf{c})$ we begin with the prover agent, which has no parents by definition. In this case, $P(\gamma_\textbf{p}) = P(\gamma_\textbf{p}|\alpha_\textbf{p})$, and we simply refer to equations \ref{eq: gamma dishonest parent} or \ref{eq: gamma honest} and pick the correct case. We then consider that parent's child agent, and to compute $P(\gamma_\textbf{c})$, we evaluate 
\[P(\gamma_\textbf{c}) = P(\gamma_\textbf{c}|\alpha_\textbf{c})P(\beta_\textbf{p})\] using equations \ref{eq: gamma dishonest child} or \ref{eq: gamma honest} accordingly. If the tree architecture has more than one depth level, the child agent then becomes the parent, and the process is repeated for its children until the leaves of the tree are reached.

It can be observed that $P(\gamma_\textbf{c})$ is therefore a binary coefficient in Equation \ref{eq-m_ij}. We evaluate this coefficient for every possible parent to child state combinations in equations \ref{eq-logic-equations}.


\begin{subequations} \label{eq-logic-equations}
    \begin{align}
    \label{eq1-logic}
        &P[\text{child} \in \overline{H} \cap T | \text{parent} \in H \cap C \cap T] = 0 \\
        &P[\text{child} \in \overline{H} \cap \overline{T}) | \text{parent} \in H \cap C \cap T) = 1 \\
        &P[\text{child} \in \overline{H} \cap T | \text{parent} \in H \cap \overline{C} \cap T] = 1 \\
        &P[\text{child} \in \overline{H} \cap \overline{T}) | \text{parent} \in H \cap \overline{C} \cap T] = 0 \\
        &P[\text{child} \in \overline{H} \cap T | (\text{parent} \in \overline{H} \cap C \cap T] = 0 \\
        &P[\text{child} \in \overline{H} \cap \overline{T} | \text{parent} \in \overline{H} \cap C \cap T] = 1 \\
        &P[\text{child} \in \overline{H} \cap T | \text{parent} \in \overline{H} \cap C \cap \overline{T}] = 0 \\
        &P[\text{child} \in \overline{H} \cap \overline{T}) | \text{parent} \in \overline{H} \cap C \cap \overline{T})] = 1 \\
        &P[\text{child} \in \overline{H} \cap T | \text{parent} \in \overline{H} \cap \overline{C} \cap T] = 0 \\
        &P[\text{child} \in \overline{H} \cap \overline{T} | \text{parent} \in \overline{H} \cap \overline{C} \cap T] = 0 \\
        &P[\text{child} \in \overline{H} \cap T) | \text{parent} \in \overline{H} \cap \overline{C} \cap \overline{T}] = 1 \\
        &P[\text{child} \in \overline{H} \cap \overline{T} | \text{parent} \in \overline{H} \cap \overline{C} \cap \overline{T}] = 0
    \end{align}
\end{subequations} 
 
\emph{Remark:} We adopt a different notation in equations \ref{eq-logic-equations} to aid with readability. For example, equation \ref{eq1-logic} is equivalent to expressing said probability as: \[P[\gamma_\textbf{c} = t, \alpha_\textbf{c} = \overline{h} | \beta_\textbf{p} = c] = 0 \] Note that only the coercion attribute of the parent affects the result of equations \ref{eq-logic-equations}. We include all three attributes to ease exposition, to show all possible state combinations.

\subsubsection{Child to parent approval}
The final stage shown in Figure \ref{fig-checks1} is whether the child will approve the parent. This approval is also entirely dependent on the state of both the parent and the child, and is summarised in Table \ref{tab:approvals}.
\begin{table*}[h]
    \centering
    \def\arraystretch{1.5}
    \begin{tabular}{|c|c|c|c|c|c|c|}
        \hline
         \diagbox[width=8em]{Child}{Parent} & $\overline{H} \cap \overline{C} \cap \overline{T}$ & $\overline{H} \cap \overline{C} \cap T$ & $H \cap \overline{C} \cap T$ & $H \cap C \cap T$ & $\overline{H} \cap C \cap \overline{T}$ & $\overline{H} \cap C \cap T$\\
        \hline
        $\overline{H} \cap \overline{C} \cap \overline{T}$  & 0 & 0 & 0 & 1 & 0 & 0 \\
        \hline
        $\overline{H} \cap \overline{C} \cap T$ & 0 & 0 & 0 & 0 & 0 & 0 \\
        \hline
        $H \cap \overline{C} \cap T$ & 0 & 0 & 1 & 1 & 0 & 0 \\
        \hline
        $H \cap C \cap T$  & 1 & 0 & 1 & 1 & 1 & 0 \\
        \hline
        $\overline{H} \cap C \cap \overline{T}$ & 0 & 0 & 0 & 1 & 1 & 0 \\
        \hline
        $\overline{H} \cap C \cap T$ & 0 & 0 & 0 & 0 & 0 & 0 \\
        \hline
    \end{tabular} 
     
    \caption{Approvals of a child to a parent.}
    \label{tab:approvals}
\end{table*} 
This table can be expressed as a matrix:
\begin{equation}
A = 
\begin{bmatrix} 
0 & 0 & 0 & 1 & 0 & 0 \\
0 & 0 & 0 & 0 & 0 & 0 \\
0 & 0 & 1 & 1 & 0 & 0 \\
1 & 0 & 1 & 1 & 1 & 0 \\
0 & 0 & 0 & 1 & 1 & 0 \\
0 & 0 & 0 & 0 & 0 & 0 \\
\end{bmatrix}
\end{equation}\label{eq-A}

Thus, all components necessary to characterise whether a parent will receive an approval from their child have been mathematically formalised. We define  $\mathcal{P}_{u,d}$ as the probability that a prover \textbf{p}, with state $x(\textbf{p}) = s_u$ will receive an approval from a witness at depth level $d$ of the witness tree. It is easy to verify that $\mathcal{P}_{u,d}$ can be expressed as follows:
\begin{equation}
    \mathcal{P}_{u,d} = e_u^T(M \odot A)^d \bold{1}
\end{equation}

where $e_u$ is the $u$-th element of the canonical base of $\mathbb{R}^6$ and $\odot$ represents the element-wise matrix multiplication (Hadamard product) and $\bold{1} = [1, 1, 1, 1, 1, 1]^T$. Then, the probability that a prover \textbf{p}, with state $x(\textbf{p}) = s_u$ will have enough approvals is equivalent to the probability that, for each level $d$ of the tree,  $D_d \geq  t\cdot n_d$, where  $D_d$ is the number of depth level approvals at a given depth level $d$. The number of approval $D_d$ can be modeled as a Binomial distribution with parameter $ \mathcal{P}_{u,d}$. Therefore, we can formalize the probability of satisfying criterion 1 as
\begin{equation}
     P(\text{Criteria $1$ $=$ True} |  s_u,\theta) =  
    P(D_d \geq  t \cdot n_d, \forall d \in \{1, ..., h\}) 
    \nonumber
\end{equation}

Which in turn can be expressed as:

\begin{equation}
\label{eq-tpopprob}
     \prod_{d = 1}^h \sum_{k =\lceil t n_d\rceil}^{n_d}  \binom{n_d}{k} \mathcal{P}_{u,d}^k (1-\mathcal{P}_{u,d})^{n_d-k}.
\end{equation}

\begin{figure}
  \centering
  \subfloat{\includegraphics[width=0.5\textwidth]{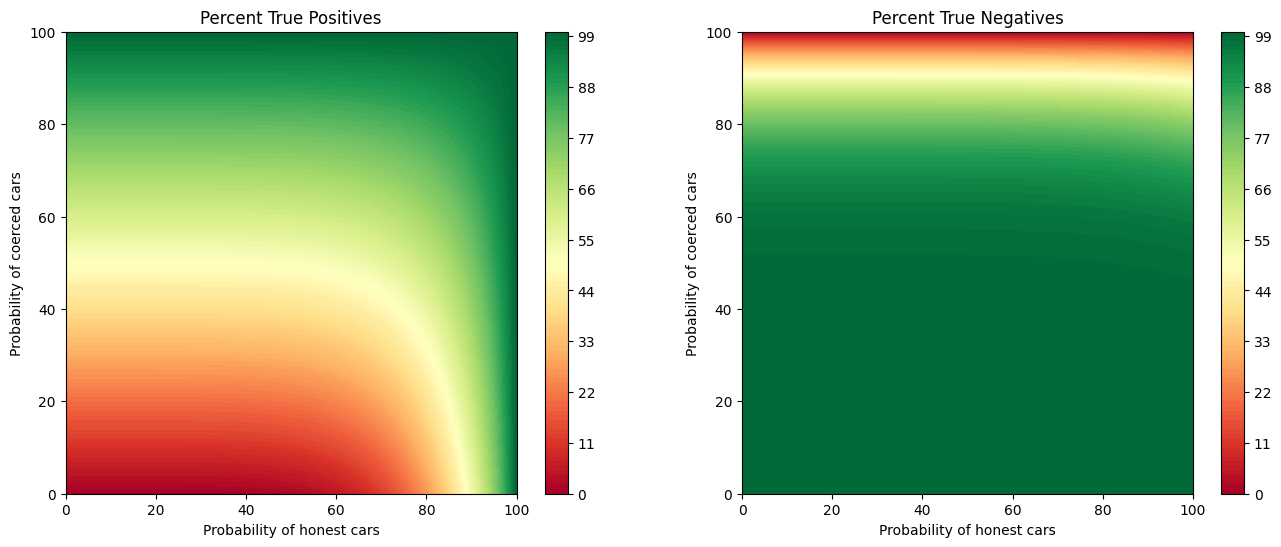}
  }\\
  \subfloat{\includegraphics[width=0.5\textwidth]{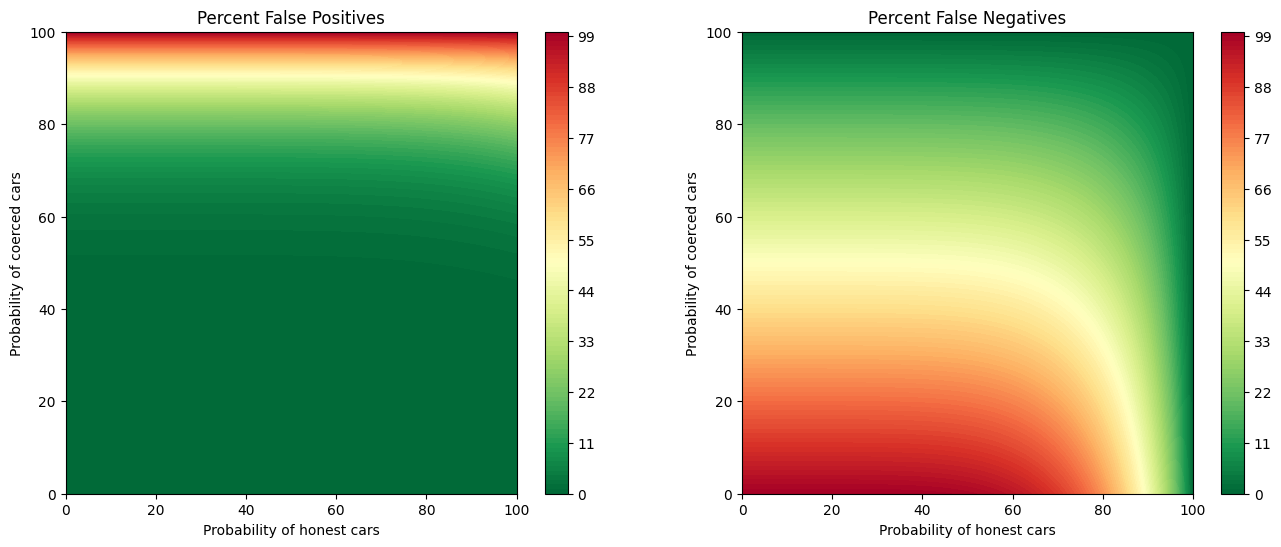}} 
  \caption{Theoretical performance of TPoP with $\theta = \{t =1, h = 1, w_1 = 6 \}$.} \label{fig:16}
\end{figure}

\begin{figure}
  \centering
  \subfloat{\includegraphics[width=0.5\textwidth]{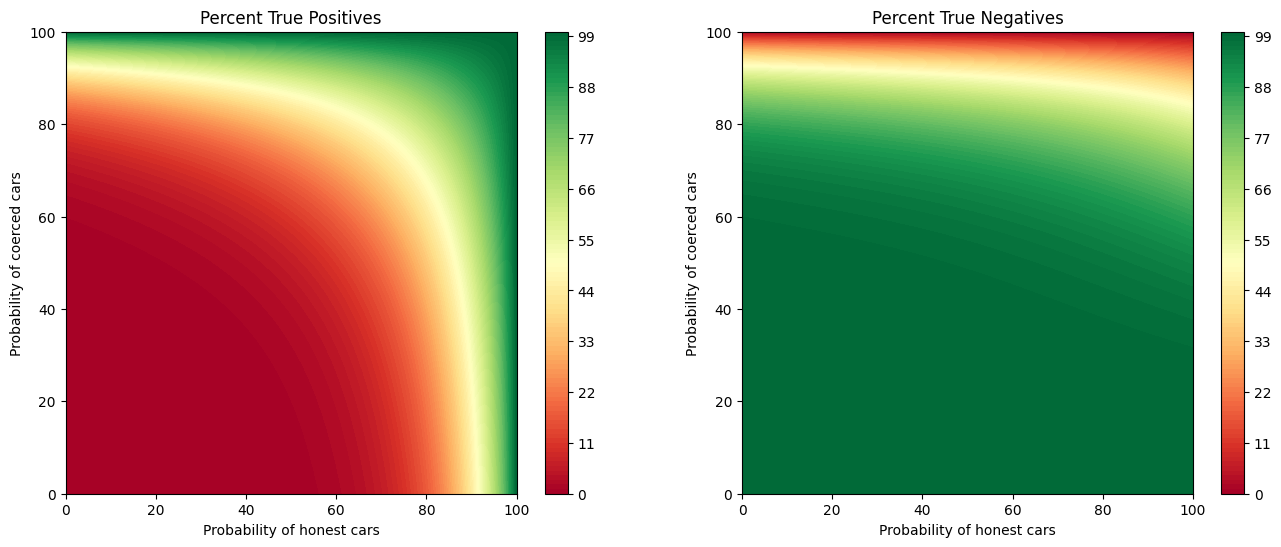}
  }\\
  \subfloat{\includegraphics[width=0.5\textwidth]{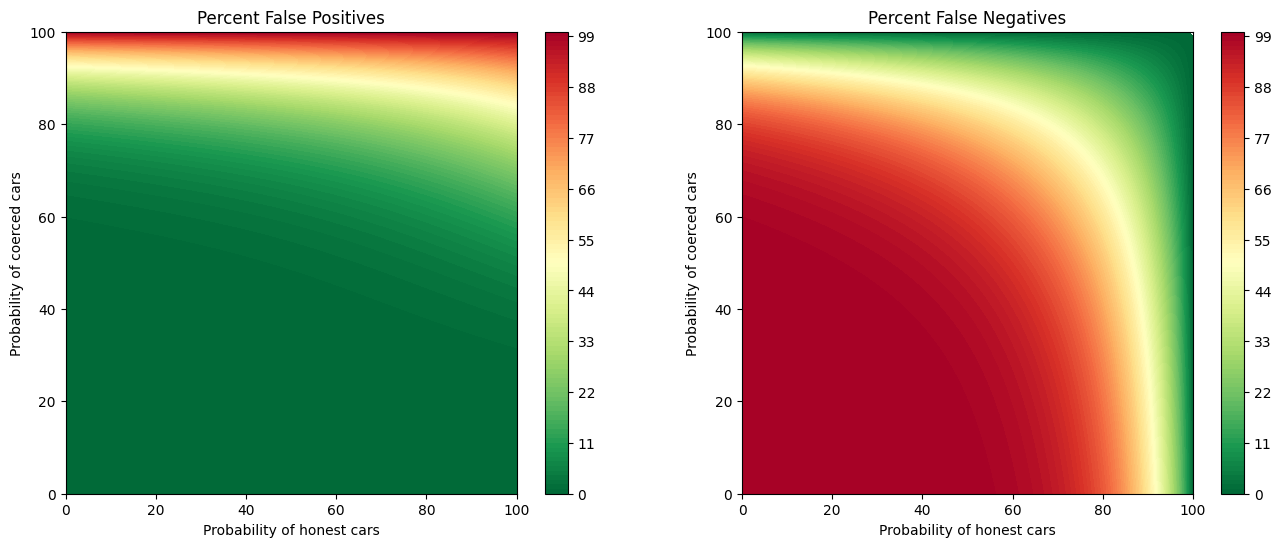}} 
  \caption{Theoretical performance of TPoP with $\theta = \{t =1, h = 2, w_1 = 2, w_2 = 2\}$.} \label{fig:22}
\end{figure}

Equation \ref{eq-tpopprob} allows us to quickly simulate the behaviour of T-PoP for different values of $\theta$ (under the assumption of an infinite density of agents). As an example, consider Figures \ref{fig:16} and \ref{fig:22} which show, for values of $p_h$ and $p_c$ ranging in the interval $[0,1]$, the performance of the T-PoP for two set of parameters, respectively: $\theta = \{t =1, h = 1, w_1 = 6 \}$ and $\theta = \{t =1, h = 2, w_1 = 2, w_2 = 2\}$. Later, in Section \ref{sec:sim} we show that the theoretical behavior matches perfectly the behaviour of a detailed agent-based simulator of TPoP.

Furthermore, we can define the probability that a child will confirm that they see the parent that selected them (i.e., that an edge exists between a parent and child). This probability can be expressed as follows:
\begin{equation}
    I_{d+1} = I_{d}(M \odot A)
\end{equation}
where $I_d$ is the  probability vector of a parent being in any given state, at depth level $d$ of the tree. Using the initial starting state in equation \ref{eq-initial state}, we can then compute the expected number of edges, $\mathbb{E}(e)$, in a tree of any given height, $h$, and branching factor, $w_d$:
\begin{align} \label{eq-edges}
    &\Vec{v} = \sum_{d = 1}^h d \cdot n_d (I_{d-1}(M \odot A)^d) \\
    \label{eq-expected edges}
    &\mathbb{E}(e) = \mathbf{1}^\intercal \Vec{v}
\end{align}

\begin{figure}
  \centering
  \subfloat[1][Tree of height 2 and branching factor 2, with threshold = 100\%.]{\includegraphics[width=0.33\textwidth]{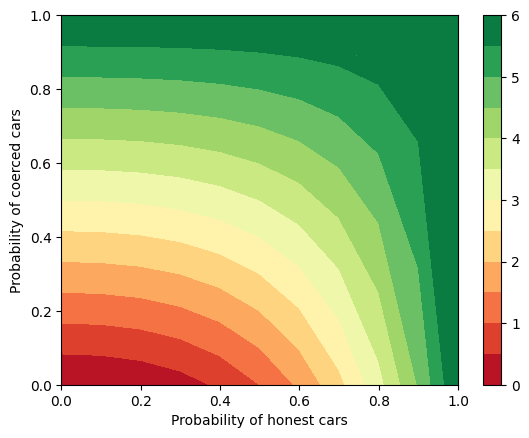}
  }\\
  \subfloat[2][Tree of height 1 and branching factor 6, with threshold = 100\%.]{\includegraphics[width=0.33\textwidth]{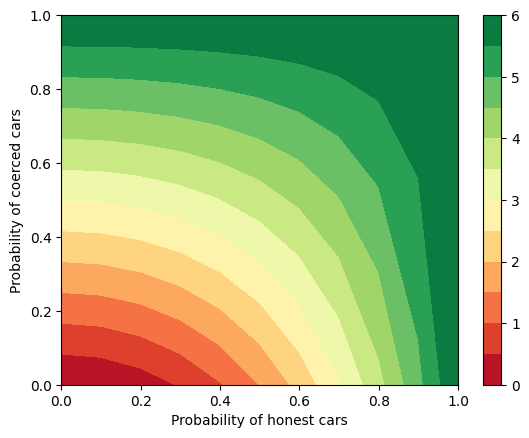}} 
  \caption{Expected number of edges in the tree.} \label{fig:edges}
\end{figure}

Using equations \ref{eq-tpopprob} and \ref{eq-edges}, we can compute the  expected number of edges in a prover's tree as the values of $p_h$ and $p_c$ range from 0 to 1. Figure \ref{fig:edges} shows the results of equation \ref{eq-expected edges}. 

\begin{figure}
  \centering
  \subfloat[1][Tree of height 2 and branching factor 2, with threshold = 100\%.]{\includegraphics[width=0.33\textwidth]{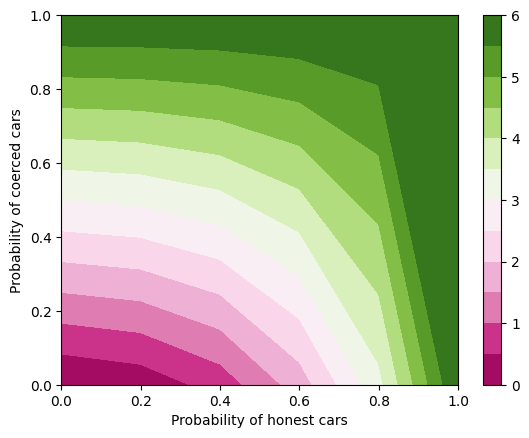}
  }\\
  \subfloat[2][Tree of height 1 and branching factor 6, with threshold = 100\%.]{\includegraphics[width=0.33\textwidth]{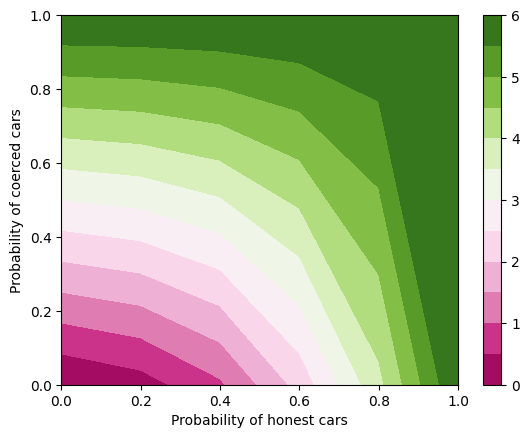}} 
  \caption{Number of edges in agent's trees after running T-PoP.} \label{fig:edges agent simulation}

\end{figure}

We then conduct 423,500 Monte-Carlo agent-based simulations, shown in Figure \ref{fig:edges agent simulation}, that count the number of edges in each agent's tree after running T-PoP. In them, we create a uniform grid of 5 by 5 square units, and a total of 3500 agents, meaning the density is 140 agents per unit square. Each agent is set to have a range of sight equivalent to the size of the unit square, meaning there are 140 agents within their range of sight. If the density of agents is not sufficiently high, an agent will not have enough edges in their tree because it does not have sufficient agents in its vicinity to name as children. We therefore purposefully select a high value of agents, to ensure that the probability of a tree not having sufficient edges is not due to a lack of nearby agents being available, but rather because the agents do not approve each other. This is precisely what we are modelling in equation \ref{eq-expected edges}.
These simulations are conducted with a threshold 100\%. The results obtained are in complete agreement with the mathematical model presented, thereby demonstrating its validity.

\subsection{Condition 2: The tree has sufficient nodes}
The second condition is modeled by considering the following assumptions:

\begin{itemize}
        \item We assume a constant average density of agents $\mu$;
    \item We assume that agents are scattered uniformly across the environment.
\end{itemize}
Accordingly, let $N_a$ be the amount of neighbours of agent $a$. This random variable can be approximated by a Poisson Point Process \cite{streit2010poisson} of parameter $\lambda = \mu \cdot \pi r^2$, where $r$ is the range of sight of the agent. It follows then that the probability that there are at least $n_d$ neighbours around agent $a$ can be immediately computed as 

\begin{equation}
    P(N_a > n_d) = 1 - P(N_a \leq n_d) = 1 - \sum_{k=0}^{n_d - 1} e^{-\lambda} \frac{\lambda^k}{k!}
    \label{eq: condition 2}
\end{equation}

This means, that for TPoP with a certain $\theta$ the probability of satisfying condition 2, $P(\text{Criteria $2$ $=$ True} |  s_u,\theta)$, for all $s_u$, can be expressed as

\begin{equation}
    \prod_{i = 0}^{h-1}\prod_{j = 1}^{n_{i}}(1 - \sum_{k=0}^{n_{i+1} - 1} e^{-\lambda} \frac{\lambda^k}{k!}),
\end{equation}

which is simply the product of the probabilities that each parent in the tree will have enough neighbours.

\subsection{Condition 3: all the nodes in the tree are unique}
In this subsection we analyze the probability of constructing a tree with distinct nodes given a certain density of agents. 
Specifically we focus on the two structures that are presented in Section \ref{sec:sim}:\newline

\begin{itemize}
    \item TPoP with $\theta_1 = (t, h = 1, w_1 = 6)$
    \item TPoP with with $\theta_2 = (t, h = 2, w_1 = 2, w_2 = 2)$.\newline
\end{itemize}

For brevity, we denote $P(\text{Criteria $3$ $=$ True} |  s_u,\theta)$ as $p(\theta)$ (as the probability of condition 3 being true does not depend on the state of the prover). In the remainder of this section we make the assumption that each agent selects witnesses among its neighbours without replacement. Although it may be an unrealistic assumption (as in practice each prover would select distinct witnesses and they would not deliberately try to hinder their own proof), it allows us to provide a lower bound on the probability of selecting a tree made of unique nodes.

Let us now assume that a population of $N$ agents are distributed uniformly at random on a plane of size $l \times l$, with every agent's range of sight set to constant $r$. We can view agents as nodes in a graph, where there exists a link between two nodes if and only if the distance between corresponding agents is less than or equal to $r$. This graph structure is called a Random Geometric Graph and it has a known mean node-degree $\mk \simeq \pi N r^2 / l^2 $ \cite{penrose2003random}. The mean node-degree $\mk$ of the graph is equivalent to the expected number of neighbours in the population of agents. 

For our calculations we make use of a function $\kappa_{n,y}$ defined as:
\begin{equation}
\label{eq:kappa}
\kappa_{n,y} = \frac{\binom{\mk-n}{y}}{\mk^y/y},
\end{equation}
representing the probability of picking $n+y$ unique agents from a set of $\mk$ neighbours, given that $n$ options are picked beforehand (eliminated) and $y$ new agents are to be picked.

For the case of parameter $\theta_1$, $p(\theta_1)$ is straightforward to compute and it is equal to $\kappa_{0,6}$.

For the tree with parameter $\theta_2$, finding the probability that all nodes are unique is more complex. Let us consider a tree started by a \emph{prover} $a_1$ at the root, tree height of $h=2$, and branching factor of $w=2$, i.e., each parent choosing two of its neighbours at random (with replacement) as its children (i.e., the nodes at level 1 will not choose themselves). 

\begin{figure}[!]
\centering
\includegraphics[width=0.48\textwidth]{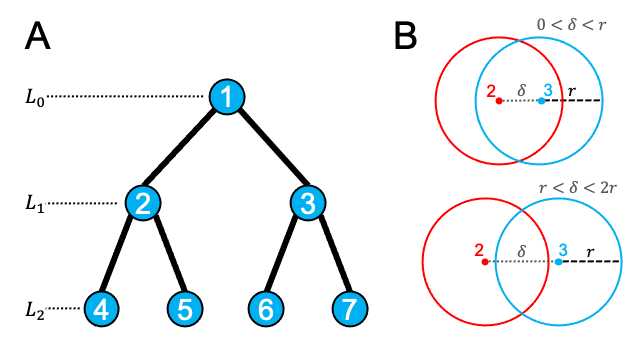}
    \caption{A) A tree of height $h=2$ and branching factor $w=2$, with agents and levels labeled. B) Two separated scenarios visualized, when agents $a_2$ and $a_3$ selected at random in the field of view of $a_1$ are $\delta<r$ nearby (top) versus $\delta>r$ far apart (bottom).}
    \label{fig:uqtree}
\end{figure}

Let $a_2$ and $a_3$ (at depth level 1 in the tree) be the children of $a_1$ (which is at level 0) in the tree, selected at random within $a_1$'s field view; $a_2,a_3\in \overline{D}_1$. Agents $a_4,\dots, a_7$ are the leaves of the tree (forming level 2); see Fig. \ref{fig:uqtree}A. We denote the probability of 7 distinct agents forming the three levels of the tree ($L_0, L_1, L_2$) as $P(L_0,L1,L2)$, which can be written as:
\begin{equation}
\label{eq:PL}
P(L_0,L_1,L_2)=P(L_0)P(L_1|L_0)P(L_2|L_0,L_1),
\end{equation}
with $P(L_0)=1$, $P(L_1|L_0)$ denoting the probability that $a_1 \neq a_2 \neq a_3$, and $P(L2|L_0,L_1)$ being the probability that $a_1,\dots,a_7$ are different agents given that $a_1,a_2,a_3$ are distinct. 

The probability of having $a_1 \neq a_2 \neq a_3$ can be calculated in a straightforward manner, since $a_1$ has an expected number of $\mk$ neighbours with none of them eliminated as options. Thus, using Eq. \eqref{eq:kappa}, we get the probability of picking two different agents at random from a set of size $\mk$, i.e., $P(L_1|L_0)=\kappa_{0,2}$. Now, with $a_2$ and $a_3$ picked at random within the field of view of $a_1$, which is a disk of radius $r$, we use the probability density function for the distance between the two agents $\delta=\|p_2-p_3\|_2$. This is simply the probability density function for the distance between two random points in a disk:
\begin{equation}
    f(\delta)=\frac{4\delta}{\pi r^2}cos^{-1}(\frac{\delta}{2r})-\frac{2\delta^2}{\pi r^3}\sqrt{1-\frac{\delta^2}{4r^2}}.
\end{equation}
If $\delta<r$ then $a_1$ is a neighbour of $a_2$ and vice versa, thus for each agent, one neighbour out of the expected number of neighbours is eliminated (in order to have distinct agents in the tree). See Fig. \ref{fig:uqtree}B (top panel). Also, for all possible values of $\delta$, i.e., $0<\delta<2r$, the probability of selecting a common neighbour by either $a_1$ or $a_2$ equals the fraction of their field of view that is overlapping. This can be calculated by $\adel$, a function of $\delta$, defined as below:
\begin{equation}
\adel=
\left[2r^2cos^{-1}(\frac{\delta}{2r})-\frac{\delta}{2}\sqrt{4r^2-\delta^2}\right]
/
\pi r^2,
\end{equation}
where the numerator of the RHS is the overlapping area of two circles with radius $r$ with centers $\delta$ far apart, and the denominator is the area of the circle with radius $r$. 

Without loss of generality, assume that $a_2$ picks two random neighbours $a_4$ and $a_5$ as its children (starting with $a_1 \neq a_2 \neq a_3$). The Prover agent $a_1$ is certainly a neighbor of $a_2$, and if $\delta<r$ ($\delta>r$) then $a_3$ is (is not) a neighbor of $a_2$, and $a_1,\dots,a_5$ are distinct with probability $\kappa_2$ ($\kappa_1$). Also, with probability $\adel$, $a_4$ is in the intersection between the fields of view of $a_2$ and $a_3$ and thus reducing the number neighbours from which $a_3$ can pick (for its children to be distinct from the agents already appearing in the tree). With probability $\adel^2$, both $a_4$ and $a_5$ are neighbours of $a_3$ and reducing $a_3$'s options by two, and with probability $\adel (1-\adel)$, $a_4$ is a neighbour of $a_3$ and $a_5$ is not. Using the above logic we can approximate the probability of having 7 distinct agents in the tree, given that $a_1 \neq a_2 \neq a_3$, i.e.:
\begin{equation}
\label{eq:PL2}
\begin{split}
P(L2&|L_0,L_1)=\\
\int_{0}^{r} &f(\delta)\kappa_{2,2} \\ &\cdot\left[(1-\adel)^2\kappa_{2,2}+2(1-\adel)\adel\kappa_{3,2}+\adel^2\kappa_{4,2}\right] d\delta \\
+\int_{r}^{2r} &f(\delta)\kappa_{1,2}\\
&\cdot\left[(1-\adel)^2\kappa_{1,2}+2(1-\adel)\adel\kappa_{2,2}+\adel^2\kappa_{3,2}\right]d\delta.
\end{split}
\end{equation}
Finally, inserting equation \eqref{eq:PL2} into equation \eqref{eq:PL}, we have the theoretical approximation for the probability of the tree being comprised of strictly distinct agents, i.e., $p(\theta_2) = P(L_0,L_1,L_2)=\kappa_{0,2}\cdot P(L2|L_0,L_1)$. This value of $P(L_0,L_1,L_2)$, given a constant range of sight $r$, is only dependent on the density of agents $N/l^2$ and can be calculated numerically. 

\begin{figure}[!]
\centering
\includegraphics[width=0.47\textwidth]{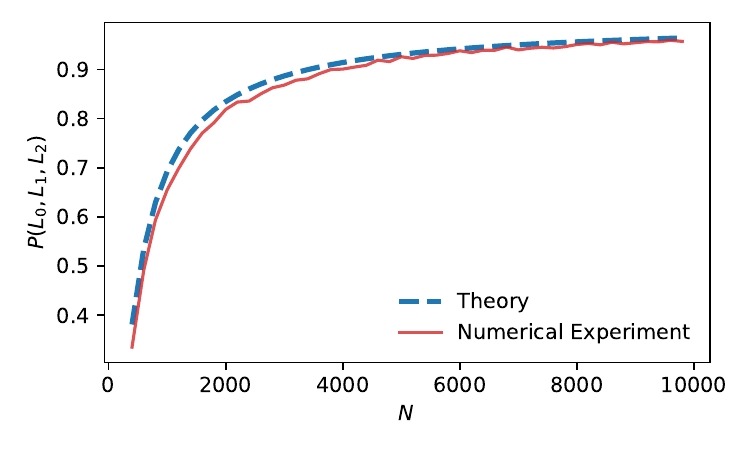}
    \caption{Validation of theoretical approximation for probability of a randomly sampled tree ($h=w=2$) containing 7 distinct agents. The results show the probability as a function of the number of agents $N$ uniformly distributed at random over a unit square plane. The red curve shows the probability calculated through a numerical experiment, where a tree is generated following the tree-building algorithm \ref{tree algorithm}, starting with a randomly selected prover. The dashed curve shows the value of the theoretical approximation denoted $P(L_0, L_1, L_2)$ calculated by equations \eqref{eq:PL}-\eqref{eq:PL2}.}
    \label{fig:validation}
\end{figure}

The theoretical approximation of $P(L_0,L_1,L_2)$ can be validated via a numerical experiment. To do so, we uniformly scatter $N$ agents on a unit 2-dimensional square, at random. A tree is then sampled, by randomly drawing a prover from the population, which then picks draws two agents at random (with replacement) within its field of view (here, $r$ is set to 0.1). We vary $N$ from 200 to 10,000, and for each $N$ sample 10,000 trees at random and count those made up of 7 distinct agents. Figure \ref{fig:validation}, compares the fraction of sampled trees with all distinct agents (red curve) with the theoretical approximation given by equations \eqref{eq:PL}-\eqref{eq:PL2}. As demonstrated by the results, the approximation performs well, converging fast to the numerical observation as the density of agents increases.

As a final remark we note that it is very difficult to to obtain a general expression to compute the probability of finding distinct agents in a tree with for any $\theta$. It is however quite straightforward to approximate $p(\theta)$, for a given $\mk$ and a given $r$, by making use of a large number of Monte-Carlo simulations.

\section{Detecting Platoon Attacks}
Given T-PoP's collaborative nature, it is susceptible to a platooning attack. In it, one \emph{dishonest} agent coerces a subset of agents to \emph{approve} their (fake) \emph{claimed} position. This cluster of agents cruise together and because the \emph{dishonest} agent has sufficient approvals, T-PoP will consistently fail in detecting that this claim is dishonest.
To address this attack vector, we construct an alternative mathematical model that defines the behaviour of these agents. With this model, it is possible to determine the likelihood that a prover tree is formed by a group of malicious platooning agents. We follow the same steps outlined in \ref{sec: math model approval}.

\subsection{Initial Agent State}
Given that a platoon is always initiated by an \emph{dishonest} agent in their \emph{claimed} position, and not their \emph{real} position, the probability of an agent being in a given state is:
\begin{subequations}
    \begin{align}
    \Tilde{I}_1 = [P(s_1), &P(s_4), P(s_5)]\\
    P(s_4) &= 0
    \end{align}
\end{subequations}
\emph{Remark:} We include state $s_4$ in the set of possible agent states, since the agent is \emph{coerced}, so it may be part of the set of agents approving the platoon \emph{prover}, however, it may never be the \emph{prover}, since it is \emph{honest}.

\begin{equation}
\Tilde{M} = 
\begin{bmatrix} 
m_{1,1} & m_{1,4} & m_{1,5} \\
m_{4,1} & m_{4,1} & m_{4,5} \\
m_{5,1} & m_{5,4} & m_{5,5} \\

\end{bmatrix}\label{eq-Mp}
\end{equation}
where the entries of $\Tilde{M}$ are computed following equation \ref{eq-m_ij}, but we only consider parent to child combinations where the agents are in states $\{s_1, s_4, s_5\}$, so $u, v \in \{1, 4, 5\}$.

Similarly, to create the platoon approvals matrix, $\Tilde{A}$, we sample the entries $a_{i,j}$, of $A$ that correspond to the approvals of all the possible agent pair combinations in states $\{s_1, s_4, s_5\}$.

\begin{equation}
\Tilde{A} = 
\begin{bmatrix} 
0 & 1 & 0 \\
1 & 1 & 1 \\
1 & 1 & 1 \\

\end{bmatrix}
\end{equation}\label{eq-Ap}

With these components, we can proceed to compute the expected number of edges in a platoon tree by substituting $\Tilde{I}_{d}$, $\Tilde{M}$ and $\Tilde{A}$ into the corresponding terms in equation \ref{eq-edges} and \ref{eq-expected edges}. 

\begin{figure}[h]
    \centering
    \includegraphics[width=0.33\textwidth]{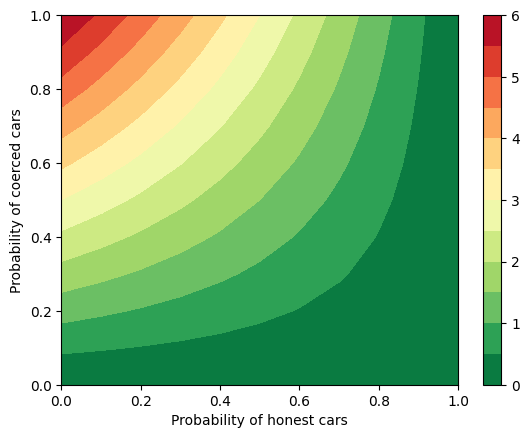}
    \caption{Expected number of edges in a platoon tree of height 1 and branching factor 6, with threshold = 100\%.}
    \label{fig:platoon t1 n6}
\end{figure}

\begin{figure}[h]
    \centering
    \includegraphics[width=0.33\textwidth]{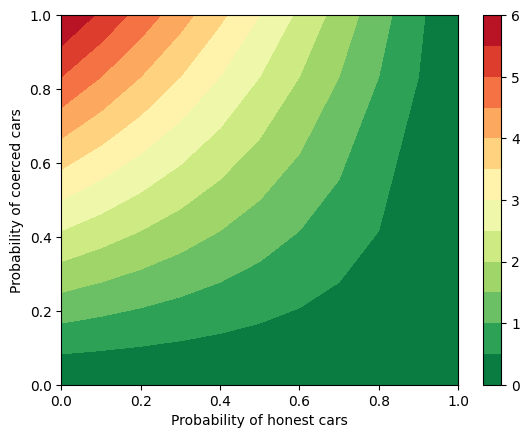}
    \caption{Expected number of edges in a platoon tree of height 2 and branching factor 2, with threshold = 100\%.}
    \label{fig:platoon t1 n2}
\end{figure}

We then show the model behaviour in Figures \ref{fig:platoon t1 n6} and \ref{fig:platoon t1 n2}. 

We highlight that to detect honest trees, it does not suffice with simply counting the number of edges. These edges must be connecting nodes that are honest, i.e., agents in states $s_3$ and $s_4$. The challenge is that the state of a node is not known or observable when a tree is assessed. However, we may find the optimal conditions such that the probability of an edge connection between agents in states $s_3$ and $s_4$ are maximised. The elements of the matrix $M$ that encode child to parent approvals that are honest are entries $m_{3,3}, m_{3,4}, m_{4,3}, m_{4,4}$. The optimal values of $p_h$ and $p_c$ that maximise the probability of these edges existing is denoted in Table \ref{optimal honest edges}.\\ 

\small
\def\arraystretch{1.25}
\begin{table}

\centering
\begin{tabular}{|c|c|c|c|}

    \hline
    $m_{i,j}$ & Parent & Child & Optimal Point\\
     \hline
    $m_{3,3}$ & $s_3$ & $s_3$ & $P(H) = 1, P(C) = 0$ \\
     \hline
    $m_{3,4}$ & $s_3$ & $s_4$ & $P(H) = 1, P(C) = 0.5$ \\
    \hline
    $m_{4,3}$ & $s_4$ & $s_3$ & $P(H) = 1, P(C) = 0.5$\\
    \hline
    $m_{4,4}$ & $s_4$ & $s_4$ & $P(H) = 1, P(C) = 1$\\
    \hline
\end{tabular}
\end{table}
\label{optimal honest edges}
\normalsize

\section{Agent Based Simulations} \label{sec:sim}
We simulate the performance of T-PoP in both high density and low density cases, under varying operating conditions. Our purpose is to determine the security and reliability guarantees that T-PoP provides in both scenarios.

\begin{figure}[ht]
  \centering
  \subfloat{\includegraphics[width=0.5\textwidth]{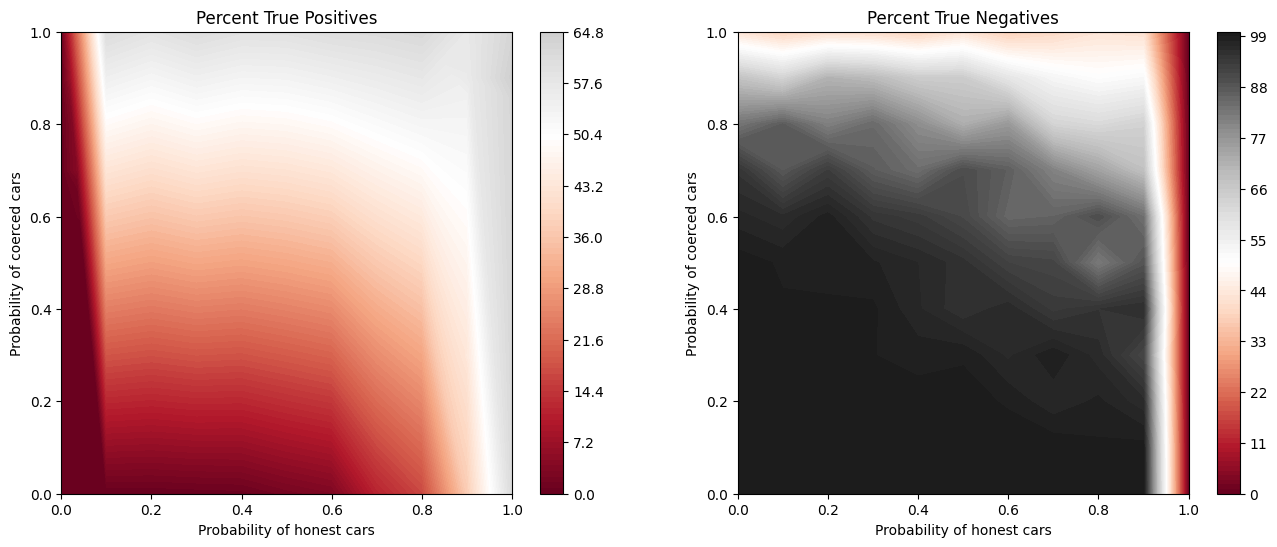}
  }\\
  \subfloat{\includegraphics[width=0.5\textwidth]{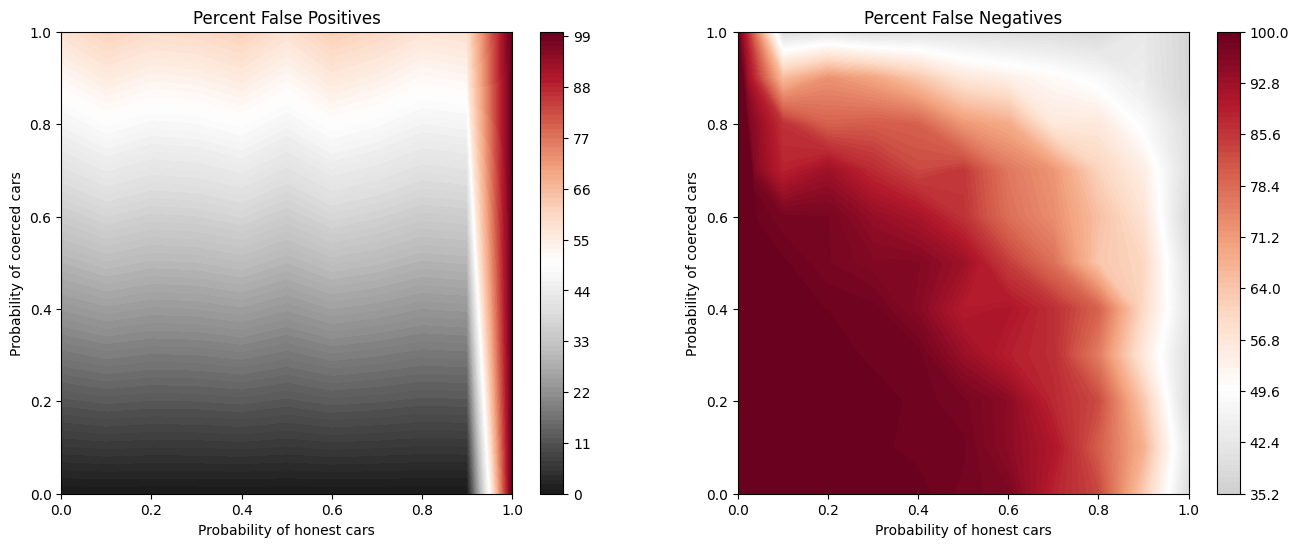}} 
  \caption{Performance of TPoP with a tree of height 2 and branching factor 2, and a threshold = 100\%.} \label{fig:t1-n2d2}
\end{figure}

\begin{figure}
  \centering
  \subfloat{\includegraphics[width=0.5\textwidth]{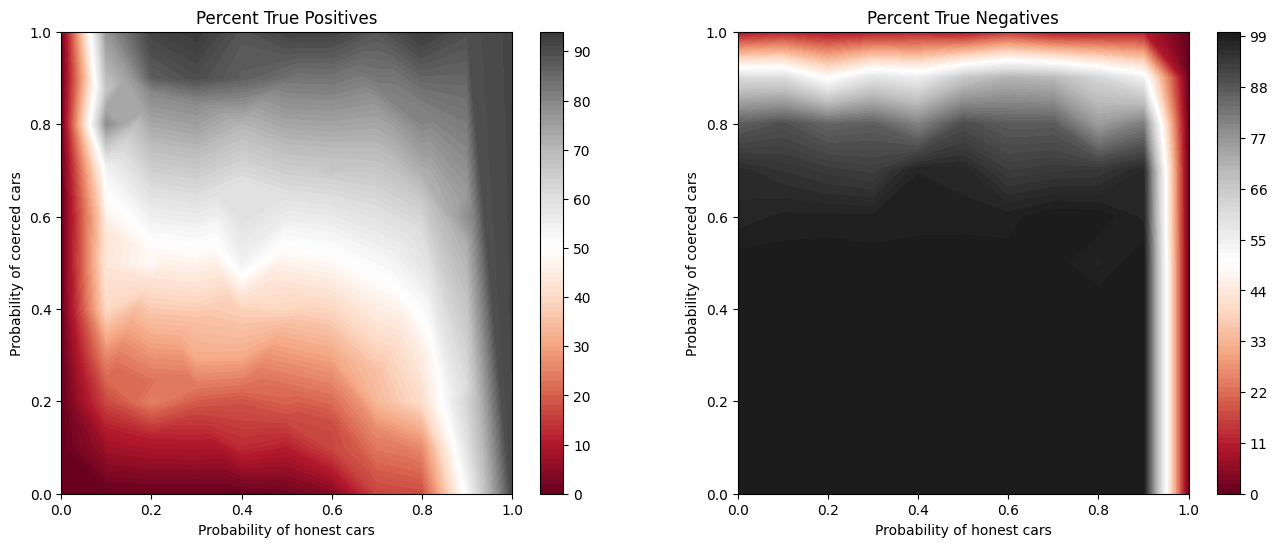}
  }\\
  \subfloat{\includegraphics[width=0.5\textwidth]{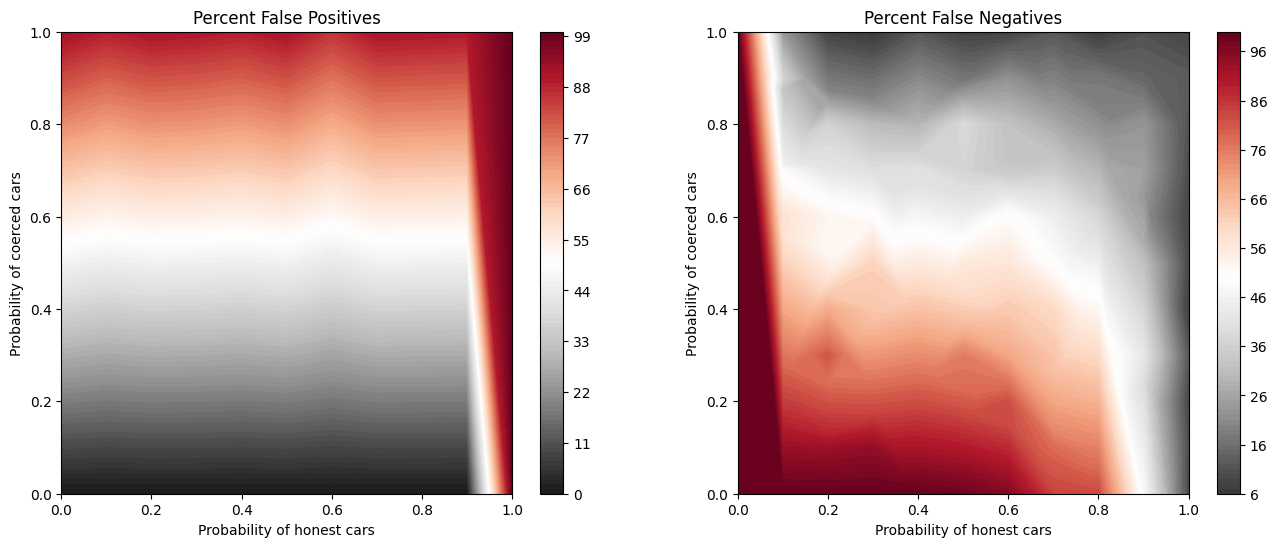}} 
  \caption{Performance of TPoP with a tree of height 1 and branching factor 6, and a threshold = 100\%.} \label{fig:t1-n6d1}
\end{figure}

\begin{figure}
  \centering
  \subfloat{\includegraphics[width=0.5\textwidth]{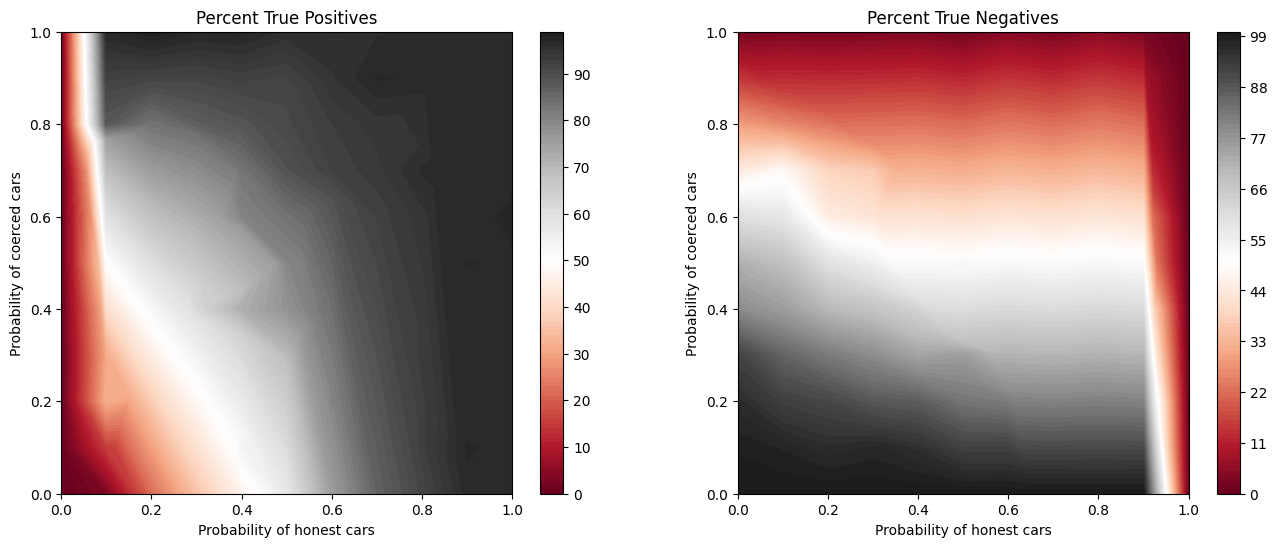}
  }\\
  \subfloat{\includegraphics[width=0.5\textwidth]{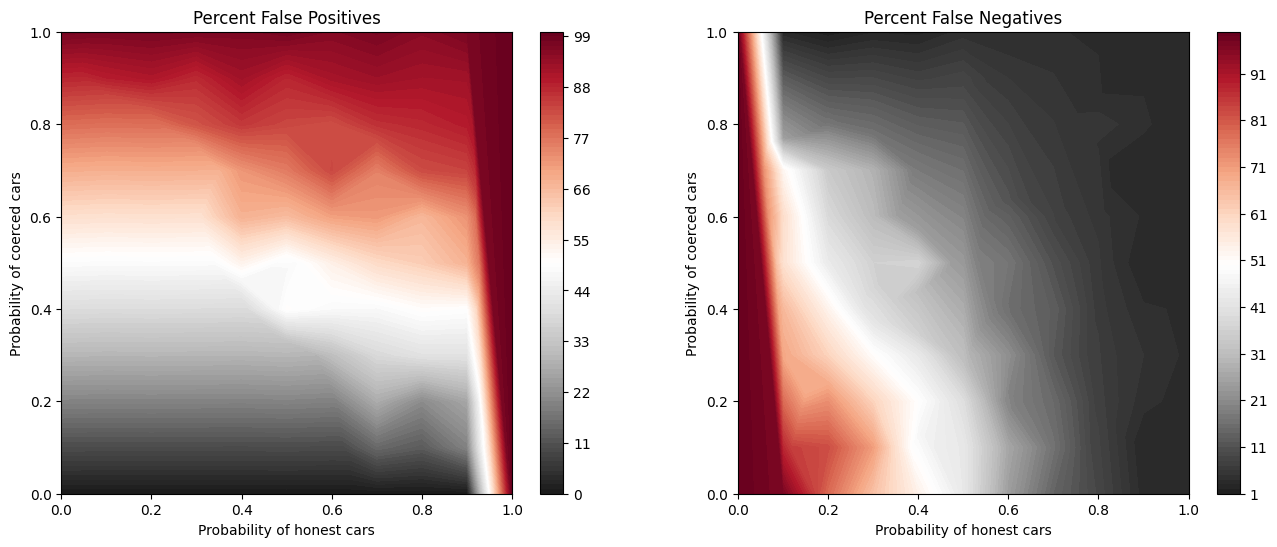}} 
  \caption{Performance of TPoP with a tree of height 2 and branching factor 2, and a threshold = 40\%.} \label{fig:t0.4-n2d2}
\end{figure}

\begin{figure}
  \centering
  \subfloat{\includegraphics[width=0.5\textwidth]{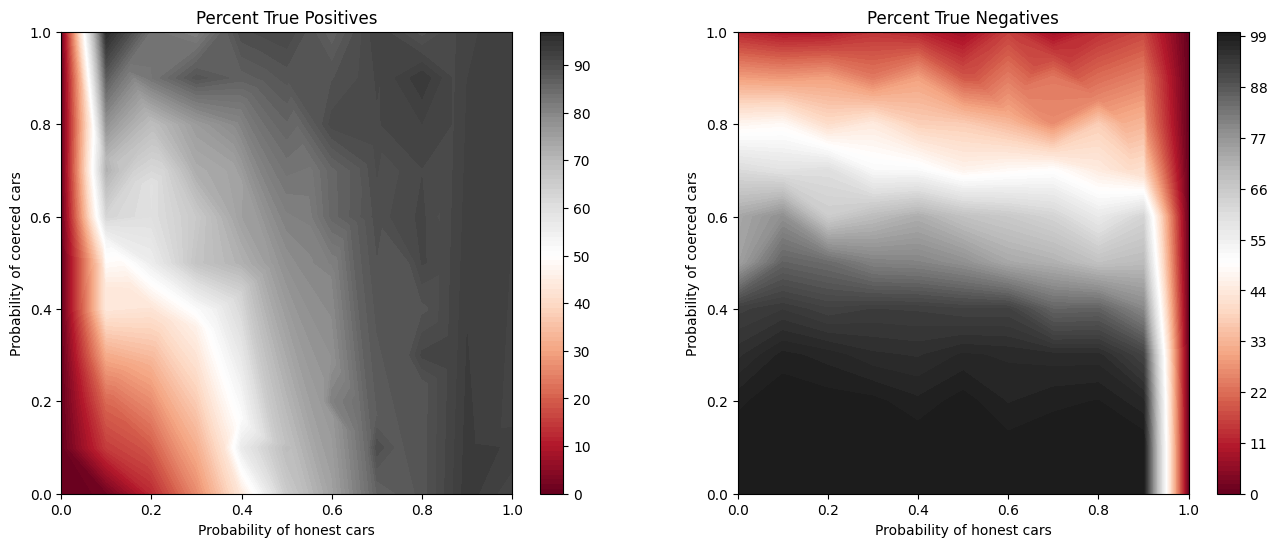}
  }\\
  \subfloat{\includegraphics[width=0.5\textwidth]{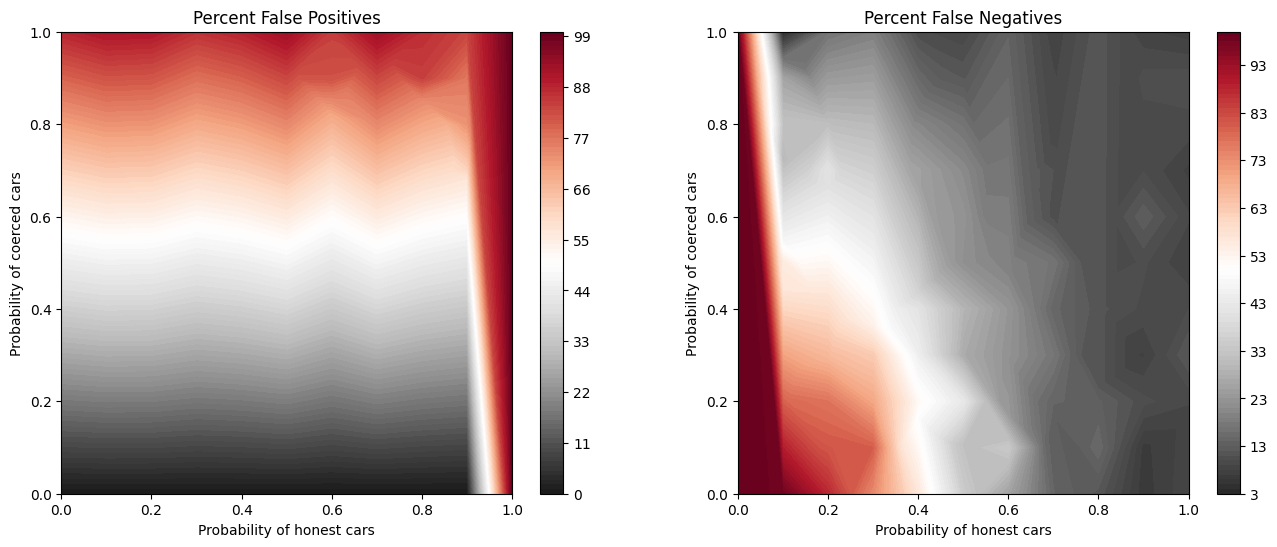}} 
  \caption{Performance of TPoP with a tree of height 1 and branching factor 6, and a threshold = 40\%.} \label{fig:t0.4-n6d1}
\end{figure}

We construct an agent-based simulator, coded in Python, which is available in our \hyperlink{https://github.com/aidamanzano/journal-tpop}{GitHub Repository}. In it, we create an environment of a given size, and instantiate a fixed number of agents with a fixed value for their range of sight. Agents are assigned an honesty, $\alpha_i$, and a coercion, $\beta_i$, attribute independently and at random, following equations \ref{eq: ph} and \ref{eq: pc}. We repeat the simulations for values of $p_h$ and $p_c$ belonging to the interval $[0,1]$.
All agents are given a real position, within the bounds of the defined environment, and dishonest agents also have a claimed position that is different to their real position. These positions are uniformly and randomly distributed across the environment.
We first simulate a low-density scenario, as shown in Figures \ref{fig:t1-n6d1}, \ref{fig:t1-n2d2}, \ref{fig:t0.4-n6d1} and \ref{fig:t0.4-n2d2}. We initialise an environment of size 10 by 10 arbitrary square units, and instantiate 350 agents, such that each has a range of sight of 1 unit. Thus, agents have an average of 3.5 agents within their field of view. Each agent constructs a tree, and then runs the T-PoP algorithms to obtain a proof-of-position. These simulations are carried out across the ranging values of $p_h$ and $p_c$, starting from 0 to 100\% in intervals of 10\%. At each combination of $p_h$ and $p_c$, 5 Monte Carlo simulations are run for every agent. In total, that is 211,750 T-PoP simulations\footnote{There are 121 possible combinations of $p_h$ and $p_c$ when varying the values at intervals of 10\%, 350 agents and 5 simulations per possible combination of $p_h$ and $p_c$ yields $5*121*350$.}.
We test the performance of T-PoP for the following operating conditions: 

   \[\theta_1 = (t = 1, w_1 = 2, w_2 = 2, h = 2)\]
    \[\theta_2 = (t = 1, w_1 = 6, h = 1)\]
   \[\theta_3 = (t = 0.4, w_1 = 2, w_2 = 2, h = 2)\]
    \[\theta_4 = (t = 0.4, w_1 = 6, h = 1)\]

In order to assess the performance of TPoP for each scenario we use an empirical approximation of the True Negatives (Security or TN) and of the True Positives (reliability ot TP). For completeness we also consider the False Negatives (FN) and the False Positives (FP). Each classification instance is described as below:
\begin{itemize}
    \item A true positive ($TP$) instance means TPoP correctly labelled an agent as honest. 
    \item A true negative ($TN$) means TPoP correctly labelled an agent as dishonest.
    \item A false positive ($FP$) means TPoP incorrectly classified a dishonest agent as honest. 
    \item A false negative ($FN$) means TPoP incorrectly classified an honest agent as dishonest.
\end{itemize}

\begin{subequations}
    \begin{align}
        TP \% &= \frac{TP}{TP + FN} \times 100 \\
        FN \% &= 100 - (TP\%)  \\
        TN \% &= \frac{TN}{TN + FP} \times 100 \\
        FP \% &= 100 - (TN\%) 
    \end{align}
\end{subequations}

Notice that we consider recall over precision because we are particularly interested in the sensitivity of the algorithm. Namely, it is of utmost importance that the number of True Negatives is high, as this value represents the security of the algorithm (i.e., how easy it is for a dishonest agent to be classified as honest).

\begin{figure}
  \centering
  \subfloat{\includegraphics[width=0.5\textwidth]{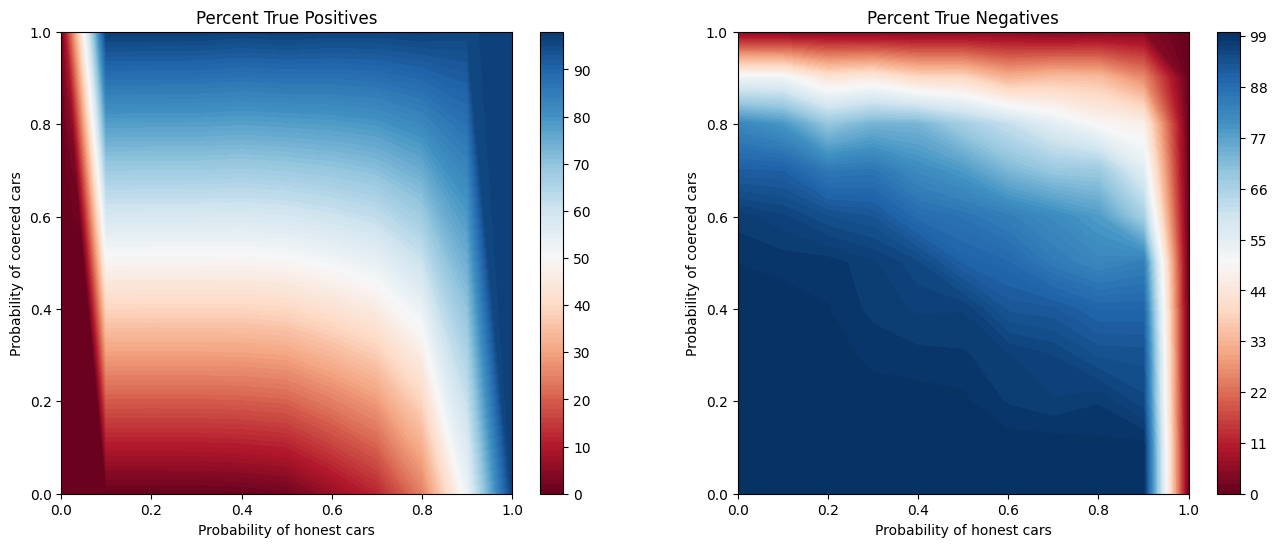}
  }\\
  \subfloat{\includegraphics[width=0.5\textwidth]{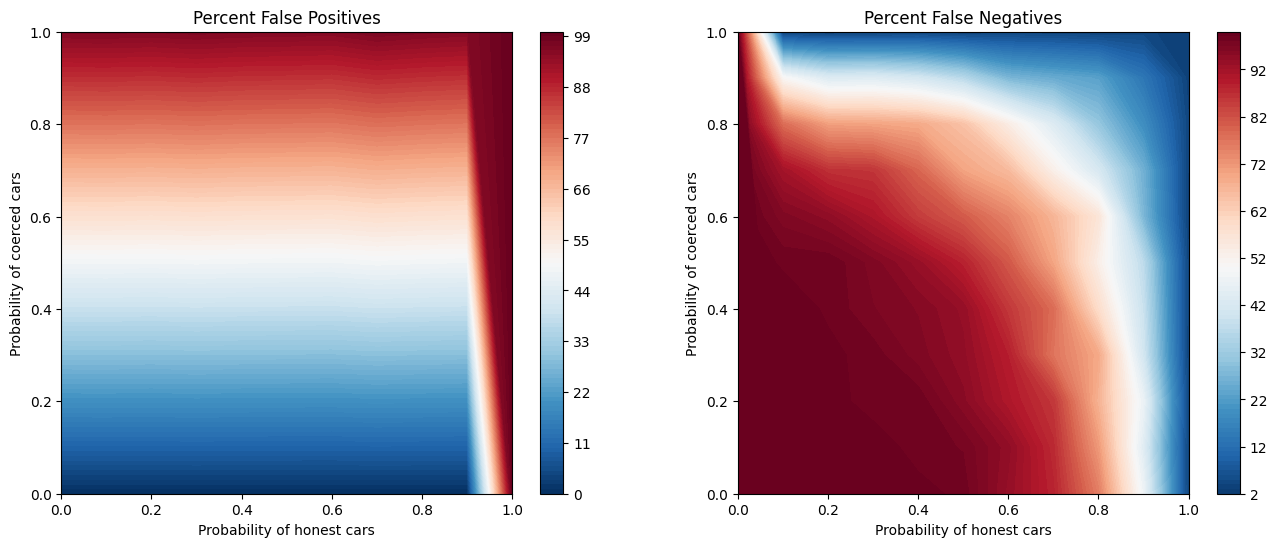}} 
  \caption{Performance of TPoP with a tree of height 2 and branching factor 2, and a threshold = 100\%.} \label{fig:t1-n2d2-1250 agents}
\end{figure}

\begin{figure}
  \centering
  \subfloat{\includegraphics[width=0.5\textwidth]{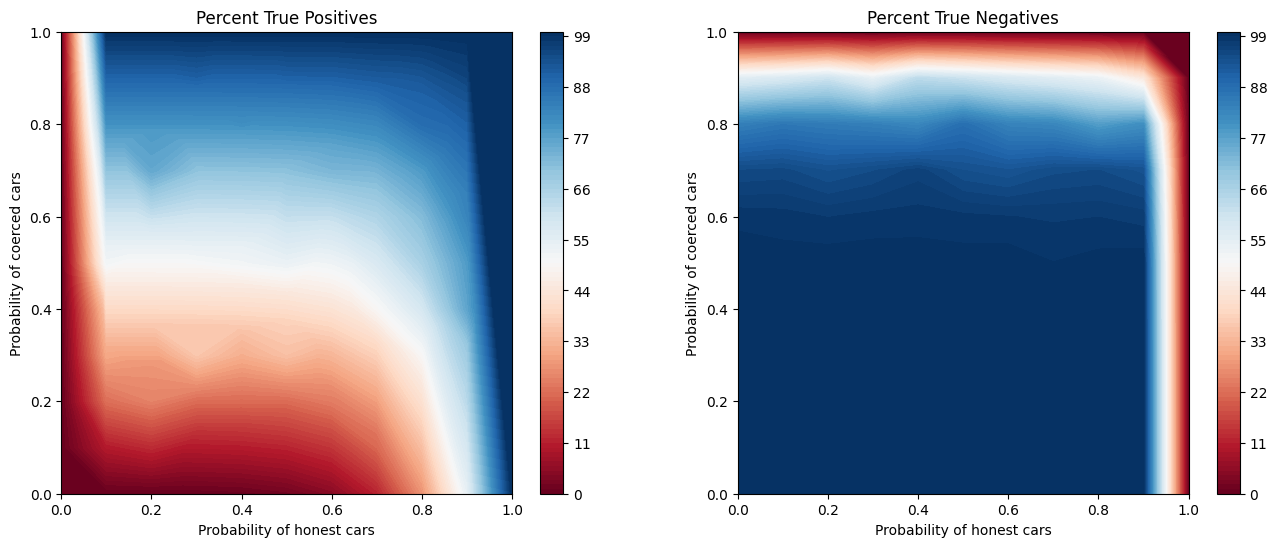}
  }\\
  \subfloat{\includegraphics[width=0.5\textwidth]{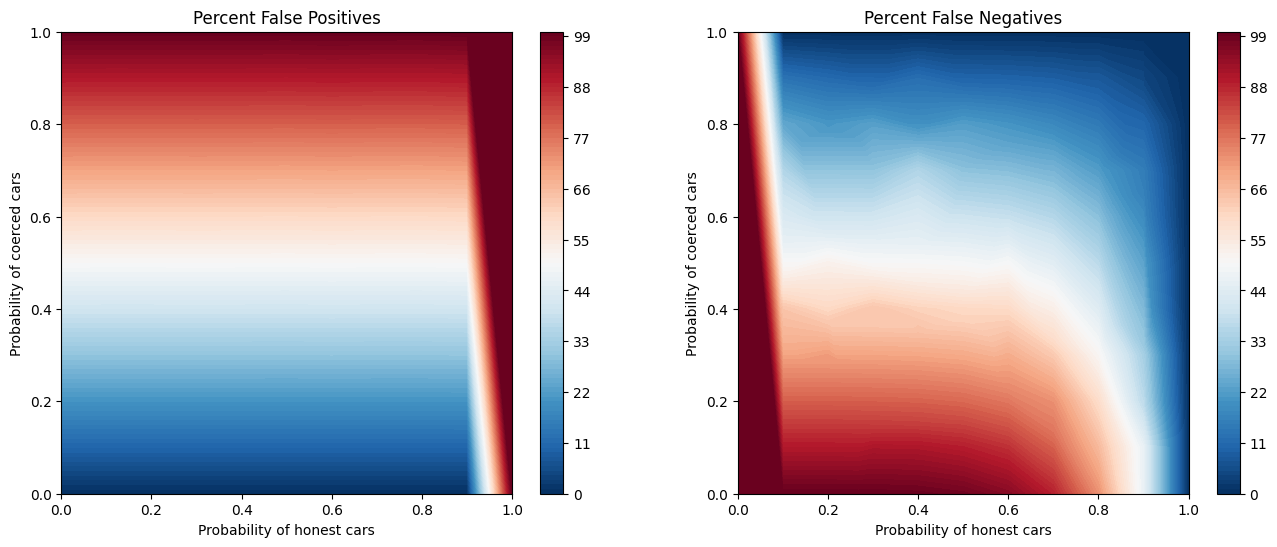}} 
  \caption{Performance of TPoP with a tree of height 1 and branching factor 6, and a threshold = 100\%.} \label{fig:t1-n6d1-1250 agents}
\end{figure}

\begin{figure}
  \centering
  \subfloat{\includegraphics[width=0.5\textwidth]{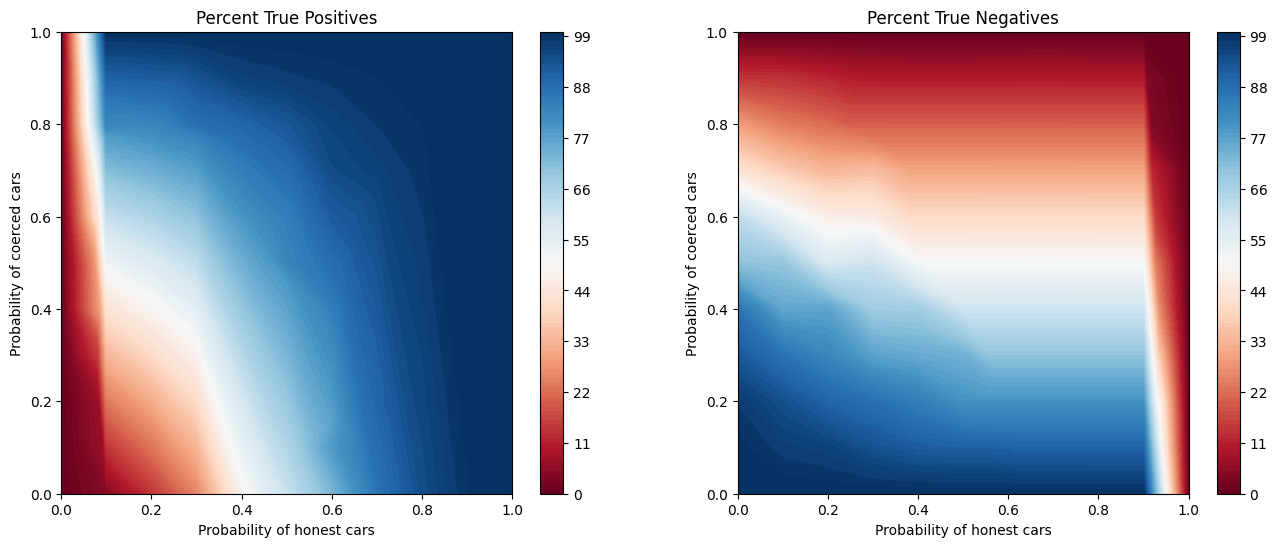}
  }\\
  \subfloat{\includegraphics[width=0.5\textwidth]{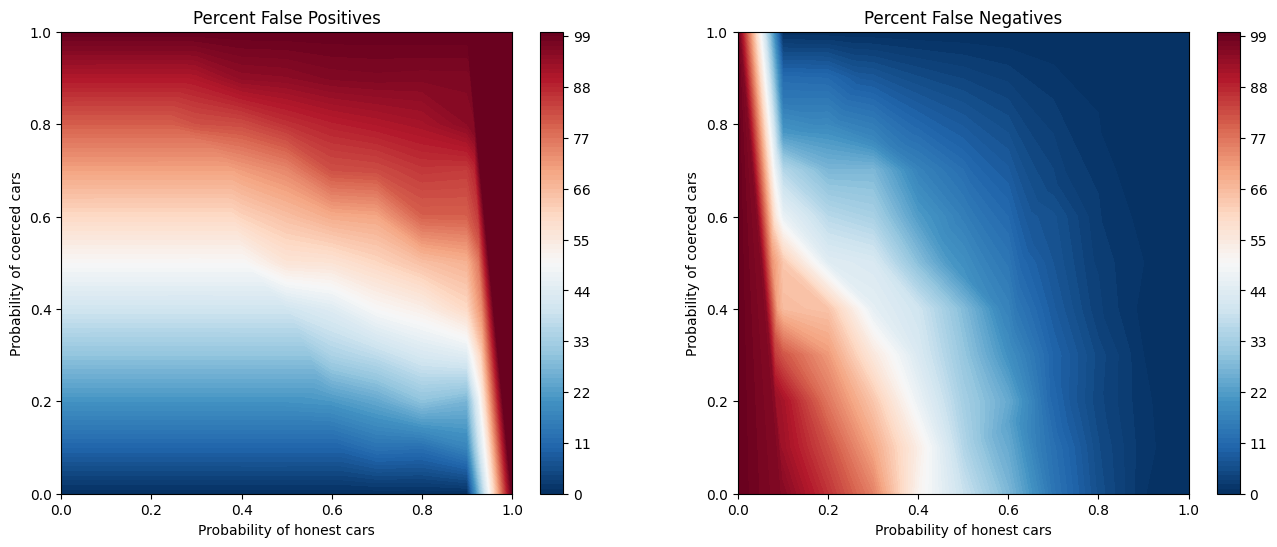}} 
  \caption{Performance of TPoP with a tree of height 2 and branching factor 2, and a threshold = 40\%.} \label{fig:t40-n2d2-1250 agents}
\end{figure}

\begin{figure}
  \centering
  \subfloat{\includegraphics[width=0.5\textwidth]{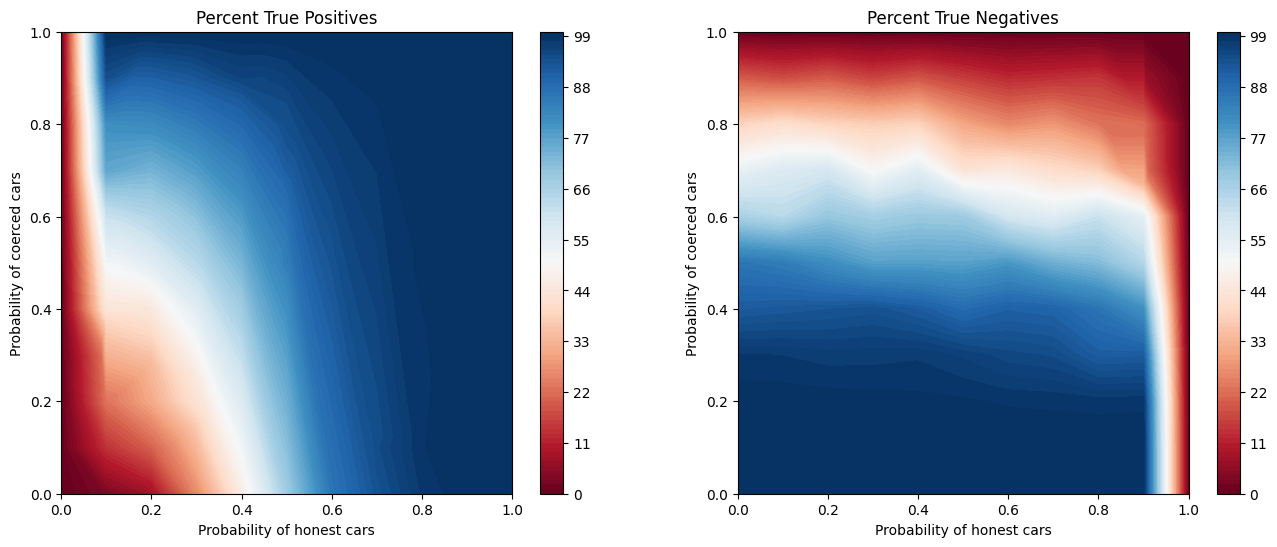}
  }\\
  \subfloat{\includegraphics[width=0.5\textwidth]{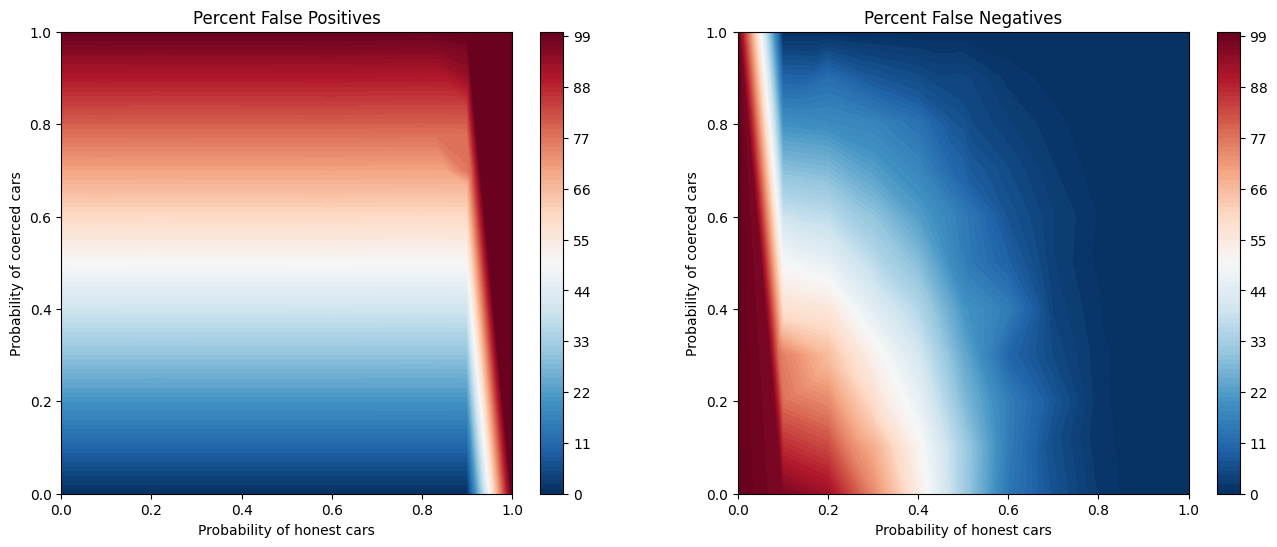}} 
  \caption{Performance of TPoP with a tree of height 1 and branching factor 6, and a threshold = 40\%.} \label{fig:t40-n6d1-1250 agents}
\end{figure}

We then consider a high-density scenario, as shown in Figures \ref{fig:t1-n6d1-1250 agents}, \ref{fig:t1-n2d2-1250 agents}, \ref{fig:t40-n6d1-1250 agents} and \ref{fig:t40-n2d2-1250 agents}. We run these simulations with the density at which the probability of selecting all unique agents in the tree tends to 1. From Figure \ref{fig:validation}, this density is found to be 50 agents per unit squared, when each agent has a range of sight of 1 square unit.  In these simulations we create a grid of 5 by 5 arbitrary square units, with 1250 agents in total, each with a range of sight of 1 square unit. We again run simulations for each possible combination of $p_h$ and $p_c$ values, at intervals of 10\%. In each combination of $p_h$ and $p_c$ values, we run T-PoP for each of the 1250 agents and repeat this process 5 times. In total, the number of simulations is 756,250. We do so for the same operating conditions previously outlined. \newline

\emph{Remark:} we note that the results of the simulations, illustrated in images 16-23, present artifacts for $p_h$ belonging to $[0,0.1]$ and $[0.9, 0.1]$. This is due to the fact that in those ranges there are either no honest agents and no honest agents (due to the resolution of $10\%$). We note however that the results, aside from these artifacts, are consistent with the theoretical results presented in Figures 8 and 9. 

\section{Performance of TPoP}

From the results obtained, it can be observed that the performance of TPoP does not vary greatly when using parameter $h = 1, w_1 = 6$ versus $h = 2, w_1 = 2, w_2 = 2$. The former slightly out-performs the latter in both thresholds tested. This is consistent with the results shown in Figures \ref{fig:edges}.
The main difference in performance is observed when varying the threshold parameter, $t$, and when varying the density. Increasing density increases the number of True Positives, given that honest agents will not be classified as dishonest for not having sufficient nodes in their tree. Decreasing the threshold improves the rate of True Positives (Reliability) of T-PoP in very low density scenarios, but this comes at the expense of Security. A lower threshold makes it easier for dishonest agents to submit a valid tree. 
As such, users may tune the threshold parameters to obtain a desired level of Security or Reliability, depending on the expected density of their system. 

\section{Conclusion}
In this paper, we provide an algorithm for agents to prove their position in a decentralised, collaborative and privacy preserving manner. The analysis focuses on the robustness of the algorithm in an adversarial environment, showing that TPoP is robust to adversarial attacks. We provide a simple mathematical model to aid users to select the most suitable operating conditions of T-PoP, provided the security and reliability guarantees they wish to achieve, and the expected agent density in their system. The mathematical model is validated through the use of extensive Monte-Carlo agent-based simulations. Our code base is open-source, and the results along with the mathematical model and the agent-based simulator can be found in our \hyperlink{https://github.com/aidamanzano/journal-tpop}{GitHub Repository}. 

\bibliographystyle{ieeetr}
\bibliography{mybib}

\end{document}